\documentclass[11pt, a4paper]{article}
\usepackage[utf8]{inputenc}
\usepackage[english]{babel}
\usepackage{amsmath,amssymb,amsfonts,amsthm}
\usepackage{geometry}
\usepackage{graphicx}
\usepackage{hyperref}
\usepackage{comment}
\usepackage{mathtools}
\usepackage{csquotes}
\usepackage{authblk}
\usepackage{physics}
\usepackage{tikz}
\usepackage{calrsfs}

\newcommand{\mc}[1]{{\color{purple}#1}}

\newtheorem{Theorem}{Theorem}
\newtheorem{proposition}{Proposition}
\begin{document}
\title{\textbf{Exact Response Theory based on Karhunen-Loève Expansions for Brownian Oscillators}}
\author[1]{M. Colangeli\footnote{ORCID:0000-0002-7424-7888}}
\author[2,3]{L. Rondoni\footnote{ORCID:0000-0002-4223-6279}}
\author[1]{P. Vozza\footnote{ORCID:0009-0000-7609-1947}}

\affil[1]{Dipartimento di Ingegneria e Scienze dell'Informazione e Matematica, Università degli Studi dell'Aquila, Via Vetoio, 67100 L'Aquila, Italy}
\affil[2]{Dipartimento di Scienze Matematiche, Politecnico di Torino, Corso Duca degli Abruzzi, 24 10129 Torino, Italy}
\affil[3]{Istituto Nazionale di Fisica Nucleare, Sezione di Torino, Via Pietro Giuria 1, 10125 Torino, Italy}

\date{\today}

\maketitle
\begin{abstract}
We extend Exact Response Theory (ERT) to an overdamped Brownian oscillator driven by thermal 
white noise. The stochastic forcing is represented by a Karhunen--Loève series and 
embedded into an autonomous, augmented phase space using a harmonic auxiliary phase.
This representation allows us to construct the ERT dissipation function for fixed mode 
amplitudes and recover stochastic averages by averaging over the amplitudes and the initial phase. 
For the variance, we derive the response at finite spectral truncation and show analytically that, 
with uniform phase sampling, the full-spectrum white-noise limit reproduces the generator-based
result for a sudden temperature change. 
Furthermore, a comparison with first-order linear response theory highlights
how this may fail in general. In particular, if the noise-induced variance  is quadratic in the noise amplitude, it is completely omitted by linear response.
The analysis provides a framework for this model and for studying more 
general forcing; extensions to non-Markovian dynamics require separate convergence and 
regularity analyses.
\end{abstract}
\section{Introduction}

Linear response theory \cite{kubo,Rphysrep} is one of the most successful frameworks in statistical mechanics, offering an intuitive approach to describe macroscopic systems near thermodynamic equilibrium in terms of their equilibrium dynamics and probability distributions \cite{Ruelle1998,ruelle2009review,mejia2025nonequilibrium}. Broadly applicable across fields such as hydrodynamics, heat and electric transport, optics \cite{Kumar20}, many-body physics  \cite{Coleman15}, and climate \cite{ghil2020physics}, it provides the foundation for classical linear transport laws \cite{kreuzer,marconi2008fluctuation}, 
which are characterized by the physically interpretable and experimentally measurable transport coefficients. 
However, although it correctly describes systems subject to driving forces that are fairly large on the scale of human activity, the theory predicates that perturbations must be weak, and that critical points and phase transitions are not allowed.
More recently, an Exact Response Theory (ERT), originating from the literature on fluctuation theorems \cite{ECM,evans2002fluctuation,evans2008fluctuation}, was developed within the field of molecular dynamics. This approach is also known as the Transient Time Correlation Function formalism \cite{todd1997application,evans2016fundamentals,maffioli}.

While the linear theory primarily targets macroscopic systems in the thermodynamic limit \cite{evans1996causality}, ERT is designed to treat systems driven far away from equilibrium, as well as small systems prone to pronounced nonequilibrium fluctuations. It has also proven effective in analyzing phase transitions \cite{amadori2022exact,DiFlorio} and a wide variety of other physical problems \cite{bernardi2016local,morriss1987application,evans1988transient}. Initially formulated for autonomous systems, the theory was later extended to periodic perturbations \cite{Petravic,Todd}, general deterministic time dependent perturbations \cite{DalCengio,Maes2020}, 
stochastic time dependent perturbations \cite{iannella2023exact}, delay equations \cite{Gollinucci},
and quantum mechanical systems \cite{greppi2025quantum}. A central quantity in this formalism is the \textit{dissipation function}, which, under suitable conditions, can be interpreted as the energy dissipation rate of the particle system, serving as the primary focus of Fluctuation Theorems \cite{evans2002fluctuation,searles2000ensemble,SRE2007}.

A remaining conceptual and practical challenge is how to construct the dissipation function when 
the forcing is a generic stochastic process: the dynamics is described by a suitable stochastic differential equation (SDE), 
whereas the usual ERT construction is formulated in 
terms of a deterministic phase-space flow and an initial density.

Generator-based methods provide a well-established tool to calculate 
the evolution of observables \cite{risken1996fokker,pavliotis2014stochastic,Maes09,Col11}. 
They constitute an appropriate benchmark here, but their formulation is inherently tied to Markov processes.
A complementary route is to represent the random forcing by a Karhunen--Loève (K--L) series and to introduce auxiliary variables that encode its time dependence as an autonomous flow. For fixed mode amplitudes, the resulting enlarged dynamics is deterministic, so that the ERT dissipation function can be constructed in the augmented phase space; averaging over the amplitudes and over the initial phase of the auxiliary variables then recovers the stochastic ensemble. 
This provides a concrete way to bring stochastic forcing into the phase-space response structure, while making explicit the averaging and limiting operations on which the construction depends. 

This representation does more than render the dynamics autonomous: it separates the forcing into
two sets of variables of different status, and the statistical treatment reserved to each
of them selects a level of description. The spectral amplitudes are constants of the augmented
motion; holding them fixed leaves a deterministic, time dependent forcing, while assigning them a
probability law turns the very same equations into a stochastic process, and neither choice affects
the phase space variation rate or the dissipation function. The initial phase of the auxiliary
oscillator, on the contrary, evolves, so that its distribution enters the dissipation function
explicitly: the uniform choice adopted in this work describes a
forcing whose time origin is not resolved, whereas a
distribution concentrated about a given phase would describe a
perturbation whose initial value is known with finite precision. Expectations may accordingly be
taken over the original phase space alone, at fixed amplitudes and fixed phase, or over the phase as
well, or over both sets of  variables; the exact response formula holds at each of these
levels, the coarser ones following from the finer ones by simple averaging. The same framework thus
encompasses deterministic as well as stochastic dynamics and interpolates between them, and it
allows one to deal with correlated as well as with white noise. Here this structure is used only in
its two extreme forms, but in Section 5 we will indicate
where it is already at work in the present calculations.

We formulate and test this approach for an overdamped Brownian oscillator, described by the spatial coordinate $q$, subjected to a sudden change in bath temperature. This model is simple enough to admit an exact 
generator-based result, yet it exposes a feature that makes the comparison informative. For the 
observable $q^2$, the change in variance caused by the noise is quadratic in its amplitude. 
Consequently, first-order response predicts only the deterministic relaxation of the initial 
variance, whereas the exact response also captures its time evolution. This is relevant, for 
instance, in the quantification of uncertainties concerning a given phenomenon \cite{Priyanka}.
The model therefore lets us assess not only whether the augmented-space calculation 
reproduces a known answer, but also which physical contribution is lost by linearization.

The paper makes four related contributions. First, we formulate ERT for the oscillator in an augmented phase space, in which the K--L representation of the noise is written in terms of Chebyshev polynomials of a single harmonic auxiliary phase. Second, we derive the exact response of the second moment in closed form for every finite truncation of the K--L series, and we characterize its convergence through an effective diffusion coefficient. Third, we prove that uniform sampling of the initial phase yields the white-noise autocorrelation in the continuum limit, and that the exact response recovers the generator-based variance formula when the number of modes diverges before the time horizon does; the two limits do not commute and must be treated explicitly. Fourth, we compare the exact response with linear response, showing that, for every observable depending on the position only, the first-order noise contributions vanish, so that linear response misses the diffusive contribution of the bath altogether.

This example establishes a precise correspondence between an augmented-space ERT calculation 
and the standard generator-based approach for one linear stochastic model and one observable.
It does not by itself establish a general theory for non-Markovian dynamics, but it 
identifies a construction whose extension to colored or memory-dependent forcing can be 
investigated by choosing a suitable stochastic representation and then checking the required 
regularity, averaging, and convergence properties.


The paper is organized as follows. Section~\ref{sec:ERT} introduces the theoretical framework and reviews the core concepts of ERT. The overdamped Langevin dynamics under study is presented in Sec.~\ref{sec:LOD}, where the mean squared displacement is computed using both a standard framework based on the forward Kolmogorov equation and the K-L expansion. Section~\ref{sec:LRT} compares the predictions of ERT against those of linear response theory for this model. Finally, conclusions are drawn in Sec.~\ref{sec:Concl}, while technical details are deferred to the Appendices.

\section{Response theory methods}
\label{sec:ERT}

In this section, we outline the theoretical foundations of ERT. To fully appreciate its scope and distinct advantages, we begin by briefly reviewing the setup of classical linear response theory, highlighting its underlying principles as well as its inherent operational limitations.

\subsection{Linear and Exact Response theory}
\label{sec:sec{LRT}}

Linear response provides a systematic framework for quantifying the effects of weak external perturbations on physical systems \cite{marconi2008fluctuation}. Following the standard approach for Hamiltonian dynamical systems, let $\mathcal{M}$ denote the phase space and consider an unperturbed system governed by the Hamiltonian $H_0(\Gamma)$, with $\Gamma\in\mathcal{M}$. At time $t=0$, an external time-dependent Hamiltonian perturbation $H_p(\Gamma,t) = -\mathcal{F}(t)\pi(\Gamma)$ is switched on, where $\mathcal{F}(t)$ represents the intensity of the perturbation, which is assumed to be small within the domain of validity of linear response,  and $\pi(\Gamma)$ is a phase variable with the dimensions of energy. The perturbed Hamiltonian is given by:
\begin{equation}
    H(\Gamma,t) = H_0(\Gamma) + H_p(\Gamma,t) = H_0(\Gamma) - \mathcal{F}(t)\pi(\Gamma).
\end{equation}
Introducing the phase space coordinates $\Gamma = (q_1, \dots, q_N, p_1, \dots, p_N)$, the perturbed equations of motion take the explicit form:
\begin{align}
    \dot{q}_j &= \frac{\partial H_0}{\partial p_j} - \mathcal{F}(t)\frac{\partial \pi}{\partial p_j} \equiv \frac{\partial H_0}{\partial p_j} - \mathcal{F}(t)k_j^q, \\
    \dot{p}_j &= -\frac{\partial H_0}{\partial q_j} + \mathcal{F}(t)\frac{\partial \pi}{\partial q_j} \equiv -\frac{\partial H_0}{\partial q_j} - \mathcal{F}(t)k_j^p,
\end{align}
where $k_j^q = \frac{\partial \pi}{\partial p_j}$ and $k_j^p = -\frac{\partial \pi}{\partial q_j}$. Denoting by $G$ the corresponding vector field, we can write:
\begin{equation}
    G(\Gamma,t) = 
    \begin{pmatrix} 
        \frac{\partial H_0}{\partial p_1} \\ 
        \vdots \\ 
        -\frac{\partial H_0}{\partial q_N} 
    \end{pmatrix} 
    - \mathcal{F}(t)
    \begin{pmatrix} 
        k_1^q \\ 
        \vdots \\ 
        k_N^p 
    \end{pmatrix} 
    = G_0(\Gamma) + G_{\text{ext}}(\Gamma,t).
\end{equation}
The time evolution of the probability density, starting from an assigned equilibrium density $f_0$, is described in terms of the Liouvillian operator $\mathcal L$, which satisfies the Liouville equation:
\begin{equation}\label{liouville_f}
    \pdv{f}{t} = - \nabla_\Gamma \cdot \left( f G \right) = -\nabla_\Gamma \cdot \left( f G_0 + f G_{\text{ext}} \right) \coloneqq -i\mathcal L f \coloneqq - i \left( \mathcal L_0+\mathcal L_{\text{ext}} \right) f,
\end{equation}
where $\mathcal L_0 \coloneqq -i \, G_0 \cdot \nabla_\Gamma$ denotes the Liouvillian operator of the unperturbed dynamics, while $\mathcal L_{\text{ext}} \coloneqq -i \, G_{\text{ext}} \cdot \nabla_\Gamma$ carries the contribution of the perturbation.
 Truncating to first order in the intensity $\mathcal F$ of the perturbation, the solution of the Liouville equation reads:
\begin{equation}
    f_t(\Gamma) = f_0(\Gamma)-i\int_0^t ds\,e^{-i(t-s)\mathcal L_0}\mathcal L_{\text{ext}}(s)f_{0}(\Gamma) + \text{H.O.},
\end{equation}
where $\text{H.O.}$ collects negligible higher-order terms in $\mathcal F$. Under this linear approximation, the time evolution of an observable $\mathcal{O}$ takes the classical form \cite{kubo,mejia2025nonequilibrium}:
\begin{equation}
    \mathbb{E}^{f_t}[\mathcal{O}] = \mathbb{E}^{f_0}[\mathcal{O}] + \int_0^t \mathrm{d}s \, R(t-s)\mathcal{F}(s),
    \label{linresponse}
\end{equation}
where the response function $R(t)$ is given by:
\begin{equation}\label{responselin}
    R(t) = \beta \mathbb{E}^{f_0}[(\mathcal{O} \circ S_0^t)\dot{\pi}],
\end{equation}
and the dissipative flux obeys $\dot{\pi} = \{\pi, H\} = \{\pi, H_0\}$, with $\{\cdot,\cdot\}$ denoting the Poisson bracket. Crucially, the subscript $0$ in $S_0^t$ designates the \textit{unperturbed} evolution, driven solely by the component $G_0$ of the full vector field $G$. Furthermore, the expectation value in Eq.~\eqref{responselin} is evaluated over the equilibrium ensemble described by $f_0$.

The fundamental result expressed by Eqs. \eqref{linresponse} and \eqref{responselin} constitutes the foundation of linear transport laws \cite{kreuzer}. Remarkably, it demonstrates that memory effects are essentially inherent to the response of dynamical systems. This fact is generally neglected when describing the behavior of thermodynamic systems because the memory decays so rapidly compared to observation times that it becomes irrelevant on the macroscopic scale. Conversely, memory effects play a crucial role in many non-thermodynamic contexts, even at the macroscopic scale \cite{mejia2025nonequilibrium}. For thermodynamic systems, this description is entirely satisfactory because relevant observables exhibit negligible fluctuations, meaning the signal observed for an individual realization practically coincides with the ensemble average. 

However, this equivalence no longer holds when non-thermodynamic (or small) systems, or strong, time-dependent perturbations, are considered. To overcome these limitations, ERT was developed to extend response theory to small systems or far-from-equilibrium regimes driven by arbitrary perturbations.
In what follows, we briefly outline the mathematical framework of ERT, that derives from works on the Fluctuation Relations, cf.\ e.g.\ Refs.\ \cite{evans2002fluctuation,SRE2007, evans2008fluctuation,Jepps16}, and enables the study of generic dynamical systems pulled far from equilibrium by perturbations of arbitrarily large amplitude, and can also treat phase transitions \cite{amadori2022exact,DiFlorio}.

We consider a generic dynamical system $S^t:\mathcal{M}\rightarrow \mathcal{M}$ on a phase space $\mathcal{M}$. Hereafter the time $t$ will be regarded as a continuous variable. We let $S^t \Gamma$ represent the solution at time $t$ of the differential equation
\begin{equation}
\dot{\Gamma}=G(\Gamma,t) = G_0(\Gamma) + G_{ext}(\Gamma,t),
\label{dyn}
\end{equation}
where $G_{ext}(\Gamma,t)$ denotes the perturbation applied to the unperturbed dynamics.
Let $f_0$ be a probability density function which evolves as prescribed by the Liouville equation such that $d\mu(\Gamma)=f_0(\Gamma)\dd \Gamma$. Performing the divergence operations in Eq.\eqref{liouville_f} and grouping terms accordingly, the Liouville equation \eqref{liouville_f} can be rewritten as:
\begin{equation}\label{liouville_omega}
    \frac{\partial f}{\partial t}(\Gamma) = \Omega^{f}(\Gamma)f(\Gamma),
\end{equation}
where $\Omega^{f_t}(\Gamma)$, called \textit{dissipation function}, is defined as
\begin{equation}
    \Omega^{f_t}(\Gamma)\coloneqq -\Lambda(\Gamma) -G(\Gamma)\cdot  \nabla_\Gamma\ln f_t(\Gamma)
\label{11}
\end{equation}
with $\Lambda(\Gamma) = \nabla_{\Gamma}\cdot G(\Gamma)$ denoting the phase space variation rate. Here we denote by $f_t$ the probability density at time $t$, obtained as the solution of Eq. \eqref{liouville_omega}
equipped with initial condition $f(\Gamma,0)=f_0(\Gamma)$.
Furthermore, we shall also denote the time integral of an observable $\mathcal{O}:\mathcal{M}\rightarrow \mathbb{R}$ along the phase space trajectory segment delimited by the points reached at time $s$ and time $t$ as $\mathcal{O}_{s,t}(\Gamma)=\int_{s}^{t}\mathcal{O}(S^r\Gamma)\, dr$.

The solution of Eq. \eqref{liouville_omega} can be expressed as: 
\begin{equation}
    f_{s+t}(S^t\Gamma) = \exp \{-\Lambda_{0,t}(\Gamma)\}f_s(\Gamma), 
    \label{eq:Lambda}
\end{equation}
or, equivalently, in the alternative form: 
\begin{equation}
    f_{s+t}(\Gamma) = \exp \{\Omega_{-t,0}^{f_s}(\Gamma)\}f_s(\Gamma),
  \label{fsol}
\end{equation}
where $\Omega^{f_0}_{s,t}(\Gamma)=\int_s^t\Omega^{f_0}(S^r\Gamma)\, d r$, which
can be written as \cite{Jepps16}: 
\begin{equation}
  \Omega_{s,t}^{f_0}(\Gamma) = \ln\frac{f_0(S^s\Gamma)}{f_0(S^t\Gamma)} - \Lambda_{s,t}.
\end{equation}
Note that while the dynamics are assumed to be invertible, this does not imply time-reversal invariance; rather, it suffices that each point along a trajectory possesses a unique pre-image under the flow.

As in the linear regime, one is primarily interested in the evolution of the observables rather than that of probability densities. The analytical derivation outlined in \cite{Jepps16} shows that the ensemble average of an observable $\mathcal{O}$ at time $t$, 
\begin{equation}
    \mathbb{E}^{f_t}[\mathcal{O}] = \int \mathcal{O}(\Gamma)f_t(\Gamma)\, d\Gamma,
\end{equation}
can be expressed as: 
\begin{equation}
\mathbb{E}^{f_t}[\mathcal{O}] = \mathbb{E}^{f_0}[\mathcal O]+\int_0^t \mathbb{E}^{f_0}[\Omega^{f_0}(\mathcal O \circ S^s)] \,d s,
 \label{ExactResponse}
\end{equation}
where $\circ$ denotes functional composition representing the evaluation of the observable along the trajectory, i.e., $(\mathcal{O} \circ S^s)(\Gamma) = \mathcal{O}(S^s\Gamma)$. Equation~\eqref{ExactResponse}, which yields the exact response formula for the observable $\mathcal{O}$, highlights that the evolution of the ensemble average of an observable can be computed entirely in terms of the initial distribution $f_0$. The key distinction from the linear response formalism is that the exact response is governed by the dissipation function $\Omega^{f_0}$, and the time evolution is evaluated along the perturbed flow $S^t$, corresponding to the full vector field $G$ in Eq. \eqref{dyn}, rather than the unperturbed flow $S_0^t$.

\subsection{Augmented phase space and K--L expansion}
\label{sec:stochper}

The framework presented in the preceding section can be extended to dynamical systems subject to rather general time-dependent forcing. The strategy draws upon the procedure established in \cite{DalCengio} and exploits auxiliary variables to enlarge the phase space and make the resulting dynamical system autonomous. Furthermore, the treatment of stochastic perturbations within the ERT framework was recently pursued in \cite{iannella2023exact} and targets SDEs of the form:
\begin{equation}
\dot{\Gamma} = G(\Gamma,t) = G_0(\Gamma) + \mathcal{F} \xi(t)
\qquad \text{for } t \in [0,T],
\label{perturbed_system}
\end{equation}
where $G$ and $G_0$ represent the perturbed and unperturbed vector fields, respectively, $\mathcal{F}$ is a constant encoding the perturbation strength, $T$ is a time constant that may be much larger than the observation timescale, and $\xi(t)$ is white noise, understood as the formal time derivative of a Wiener process $W(t)$ with the same dimension as the phase space $\mathcal{M}$.
By applying the K–-L expansion for Brownian motion, reviewed in Appendix~\ref{app:appA}, the noise term can be formally expressed as:
\begin{equation}
\xi(t) = \sum_{n=-\infty}^{+\infty} \alpha_n e^{i\beta_n t} = \sum_{n=-\infty}^{+\infty} \alpha_n \xi_n (\theta, \phi) = \xi(\theta, \phi),
\label{expansion}
\end{equation}
where $\beta_n$ denote the frequencies of the K--L modes.
Assuming the perturbation is periodic with a period much larger than any physically relevant timescale \cite{kubo}, this periodicity suggests introducing a harmonic oscillator in an auxiliary space $\mathcal{M}_{\theta\phi}$ to eliminate explicit time dependence from the governing vector field:
\begin{equation}
\label{oscillator}
\begin{cases}
\dot\theta = b\phi \\
\dot\phi = - b\theta ,
\end{cases}
\end{equation}
with $b=\pi/2 T$ recalling the frequency of the first term in the K--L expansion (see Appendix ~\ref{app:appA}). 
We stress the fact that practical computations will start from finite $T$ and finite number $N$ of modes. Then, extrapolations for the $N \to \infty$ and $T \to \infty$ limits will be attempted, keeping in mind that these limits do not necessarily commute; cf.\ below the explicitly worked out example.

The dynamics now reside in the augmented phase space $\tilde{{\mathcal M}}\coloneqq \mathcal{M}_{\Gamma} \times \mathcal M_{\theta\phi}$ whose phase space variable is $\tilde{\Gamma}=(\Gamma,\theta,\phi)$. Substituting \eqref{expansion} into \eqref{perturbed_system} and explicitly writing the components of the vector field, we obtain:
\begin{equation}
\dot{\tilde{\Gamma}} (t) = \tilde {G}(\Gamma,\theta,\phi) = \begin{pmatrix}
G_1(\Gamma) + \sum_{n=-\infty}^{+\infty} \alpha_{n1} \xi_n (\theta, \phi) \\
G_2(\Gamma) + \sum_{n=-\infty}^{+\infty} \alpha_{n2} \xi_n (\theta, \phi) \\
\vdots \\
G_d(\Gamma) + \sum_{n=-\infty}^{+\infty} \alpha_{nd} \xi_n (\theta, \phi) \\
b\phi \\
-b\theta
\end{pmatrix}
\end{equation}
where the angular variables obey Eq.~\eqref{oscillator} and the constant $\mathcal{F}$ has been absorbed into the coefficients. In this manner, ERT can be applied with minor adaptations to accommodate the time-dependent case. In the augmented space, the phase space variation rate $\tilde \Lambda = \operatorname{div} \tilde G$ equals the unperturbed rate due to the null divergence contribution of the harmonic oscillator dynamics~\eqref{oscillator}:
\begin{equation}
\tilde\Lambda(\tilde \Gamma) = \Lambda(\Gamma).
\end{equation}
Given probability distributions $f_0 (\Gamma)$ and $g_0 (\theta, \phi)$ over $\mathcal{M}_{\Gamma}$ and $\mathcal{M}_{\theta \phi}$ respectively, the unperturbed distribution over the augmented phase space $\dd \tilde \mu$ on $\tilde{\mathcal{M}}$ can be
expressed by the factorized probability density:
\begin{equation}
\tilde{f}_0 (\Gamma, \theta, \phi) = f_0 (\Gamma) g_0 (\theta, \phi).
\label{f0tilde}
\end{equation}
which is appropriate under the natural asumpion that the intial distribution in the orginal phase space and that concering the perturbation
are idependent. Then,
the dissipation function $\widetilde{\Omega}^{\tilde{f}_0}$ for the augmented phase space takes the form \cite{iannella2023exact}:
\begin{equation}
\begin{split}
\widetilde{\Omega}^{\tilde{f}_0} (\Gamma, \theta, \phi) & = -\tilde{\Lambda}(\Gamma, \theta, \phi) - \frac{\dd}{\dd t} \ln \tilde{f}_0 (\Gamma, \theta, \phi) \\
& = -\tilde{\Lambda}(\Gamma, \theta, \phi) -\frac{1}{\tilde{f}_0 (\Gamma, \theta, \phi) } \nabla \tilde{f}_0 (\Gamma, \theta, \phi)\cdot \tilde{G}(\Gamma, \theta, \phi) \\
\label{dissipation}
& = \Omega^{f_0}(\Gamma) - \frac{1}{f_0 (\Gamma)} \sum_{n=-\infty}^{+\infty} \xi_n (\theta, \phi) \sum_{k=1}^d \alpha_{nk} \frac{\partial f_0}{\partial \Gamma_k} (\Gamma) \\
&\quad - \frac{1}{g_0 (\theta, \phi)} \bigg(b \phi \frac{\partial g_0}{\partial \theta} (\theta, \phi) - b\theta \frac{\partial g_0}{\partial \phi} (\theta, \phi) \bigg).
\end{split}
\end{equation}
Let $\hat{\mathbb{E}}^{\tilde{f}}[\cdot]$ denote the expectation with respect to the density $\tilde{f}$, where the hat symbol stands for an individual realization of the noise.
Then, the exact response formula for a generic observable $\mathcal{O}$ and for a given realization of the stochastic perturbation takes the form \cite{iannella2023exact}:
\begin{equation}
\label{exact_response_augmented_starting}
\hat{\mathbb{E}}^{\tilde f_t} [\mathcal{O}] 
= \hat{\mathbb{E}}^{\tilde{f}_0}[\mathcal{O}]+\int_0^t\hat{\mathbb{E}}^{\tilde f_0} [\widetilde{\Omega}^{\tilde{f}_0}(\tilde{\mathcal{O}} \circ \tilde{S}^s)] \dd s,
\end{equation}
where $\tilde f_{t}(\tilde\Gamma) = \exp \{\widetilde{\Omega}_{-t,0}^{\tilde{f}_0}(\tilde{\Gamma})\}\tilde f_0(\tilde \Gamma)$ (see Eq.~\eqref{fsol}) represents the probability density in the augmented phase space at time $t$.
The first term on the right-hand side of \eqref{exact_response_augmented_starting} equals the expectation with respect to the unperturbed distribution, i.e., $\hat{\mathbb{E}}^{\tilde{f}_0}[\mathcal{O}] = \mathbb{E}^{f_0}[\mathcal{O}]$, while the second expectation requires evaluation over the extended phase space $\mathcal{M}_{\theta \phi}$. More precisely:
\begin{equation}
\label{response_inintegral}
\hat{\mathbb{E}}^{\tilde f_0} [\widetilde{\Omega}^{\tilde{f}_0}(\tilde{\mathcal{O}} \circ \tilde{S}^s)] =
\int_{\mathcal{M}_{\theta \phi}}
\dd \theta \dd \phi \,
g_0 (\theta, \phi)  \int_{\mathcal{M}} \dd \Gamma \,\widetilde{\Omega}^{\tilde{f}_0} (\Gamma, \theta, \phi) \mathcal{O} \big( S^s (\Gamma) \big) f_0 (\Gamma) 
\end{equation}
Since the observable $\mathcal{O}$ depends solely on $\Gamma$, its embedding in the augmented phase space does not alter its value, i.e., $\tilde{\mathcal{O}}\!\left( \tilde{S}^s (\tilde{\Gamma}) \right) = \mathcal{O} \!\left( S^s (\Gamma) \right)$, where $S^s$ and $\tilde{S}^s$ denote the perturbed flows in $\mathcal{M}_{\Gamma}$ and in $\tilde{\mathcal{M}}$, respectively. 
A mathematically convenient and physically sound choice for $g_0$
is the uniform distribution on the unit circle, as its derivative along the oscillator flow vanishes (i.e. $b \phi \partial_{\theta} g_0-b\theta \partial_{\phi}g_0=0$, which makes the last term in Eq.~\eqref{dissipation} vanish), and it also preserves the condition of ergodic consistency 
\cite{evans2002fluctuation},when it is required\footnote{Essentially, the notion that an initial probability measure $\mu_0$ and its evolved descendants $\mu_t$ are mutually absolutely continuous for every finite $t$.}.

Upon averaging over the realizations of the stochastic process, we write the exact response formula as:
\begin{equation}
\mathbb{E}^{\tilde{f}_t} [\mathcal{O}] = \mathbb{E}^{f_0}[\mathcal{O}] + \int_0^t \mathbb{E}^{\tilde{f}_0} \!\left[\widetilde{\Omega}^{\tilde{f}_0}(\tilde{\mathcal{O}} \circ \tilde{S}^s)\right] \dd s,
\label{eq:ERTst}
\end{equation}
where, within the K--L framework, $\mathbb{E}^{\tilde f}[\cdot]$ denotes a double expectation both over the noise and with respect to the distribution $\tilde f$.

The primary objective of this work is to apply this augmented formalism to an exactly solvable model, specifically an overdamped Langevin equation, thereby 
proving the equivalence of ERT and
the formalism based on the generator of the process, 
in view of the applicability of ERT to a much wider class of systems.

\subsection{Generator-based exact response}
\label{sec:gener}

We adopt the notation introduced in \cite{Vozza25} and consider the SDE for a variable $x \in \mathbb{R}^d$ formulated in Itô form:
\begin{equation}
\dd x(t) = A(x)\, \dd t + \sigma(x)\, \dd W(t), \quad x(0)=x_0, \label{pertSDE}
\end{equation}
where $A: \mathbb{R}^d \to \mathbb{R}^d$, $\sigma: \mathbb{R}^d \to \mathbb{R}^{d \times m}$, and $W(t)$ is a Wiener process in $\mathbb{R}^m$.
The generator $L$ of the process acts on state functions $\mathcal{O}: \mathbb{R}^d \to \mathbb{R}$ as follows:
\begin{equation}
L \mathcal{O}(x) = A(x) \cdot \nabla \mathcal{O}(x) + D(x) : \left( \nabla \nabla \mathcal{O}(x) \right), \label{gen}
\end{equation}
where $D(x) = \frac{1}{2} \sigma(x) \sigma(x)^T$ is the diffusion matrix, and $\nabla \nabla \mathcal{O}$ denotes the Hessian matrix of $\mathcal{O}$.

The $L^2$-adjoint of the generator $L$ acts on probability densities $f_t(x)$ according to \cite{risken1996fokker, balakrishnan2020elements}:
\begin{equation}
L^* f_t(x) = \nabla \cdot \left[ -A(x) f_t(x) + \nabla \cdot \left( D(x) f_t(x) \right) \right].
\label{adj}
\end{equation}
The corresponding forward Kolmogorov equation (or Fokker–Planck equation) for $f_t(x)$ can be written as:
\begin{equation}
\partial_t f_t(x) = \Omega^{f_t}(x)\, f_t(x)
\label{forwop}
\end{equation}
where the state function $\Omega^{f_t}(x) := (L^* f_t(x))/f_t(x)$ (defined on the support of $f_t(x)$) inherits, for the stochastic dynamics expressed by Eq.~\eqref{pertSDE}, the role of the dissipation function introduced in Sec.~\ref{sec:sec{LRT}} for deterministic dynamics obeying the Liouville equation. This quantity attains the explicit form:
\begin{align}
\Omega^{f_t}(x) 
&= -\nabla \cdot A(x) 
- A(x) \cdot \nabla \log f_t(x) 
+ \nabla \nabla : D(x) \notag \\
&\quad + 2\, (\nabla \cdot D(x)) \cdot \nabla \log f_t(x) \notag \\
&\quad + D(x) : \left[ \nabla \nabla \log f_t(x) + \nabla \log f_t(x) \otimes \nabla \log f_t(x) \right].
\label{Om}
\end{align}
The generator-based ERT for dynamics governed by Eq.~\eqref{pertSDE} can thus be expressed as \cite{Vozza25}:
\begin{equation}
\mathbb{E}^{f_t}[\mathcal{O}] = \mathbb{E}^{f_0}[\mathcal{O}] + \int_0^t \mathbb{E}^{f_0}[\Omega^{f_0}(x(0)) \mathcal{O}(x(s))] \dd s,
\label{ERTgen}
\end{equation}
which matches the structure of Eq.~\eqref{ExactResponse}. 

\section{Brownian Oscillators}
\label{sec:LOD}

In this section, we consider a paradigmatic model of dissipative processes, namely the overdamped Langevin dynamics of a Brownian oscillator~\cite{risken1996fokker, balakrishnan2020elements, colangeli2025hybrid}, for which ERT can be explicitly computed via both the K--L expansion and generator-based approaches.
We first discuss the application of ERT to these dynamics and then investigate analytical truncations of the K--L expansion.

\subsection{Exact response formula}
\label{sec:ERform}

We consider the SDE governing the motion of a particle in a harmonic potential for $t \ge 0$:
\begin{equation}
\dd q(t) = -\frac{\omega^2}{\gamma} q(t) \, \dd t + \sqrt{2D} \, \dd W(t), \quad q(0) = q_0,
\label{eq:langevin}
\end{equation}
where $W(t)$ is a standard Wiener process, $\gamma$ is the friction coefficient per unit mass, $\omega$ is the natural frequency, and $D = (m \beta \gamma)^{-1}$ is the diffusion coefficient, with $m$ being the mass of the particle and $\beta$ the inverse temperature of the bath.

We assume that the initial state $q(0)$ is drawn from a Gaussian distribution $f_0(q)$:
\begin{equation}
f_0(q) = \sqrt{\frac{m \beta_0 \omega^2}{2\pi}} \exp\left(-\frac{1}{2} m \beta_0 \omega^2 q^2\right).
\label{eq:initial_pdf}
\end{equation}
The perturbation, introduced at $t = 0$, consists of a sudden change in the temperature of the thermal bath, shifting the inverse temperature abruptly from $\beta_0$ to $\beta$.

A straightforward application of Eq.~\eqref{ERTgen} to the model described by Eq.~\eqref{eq:langevin} yields the generator-based ERT for the observable $\mathcal{O}=q^2$ in the form \cite{Vozza25}:
\begin{equation}
\mathbb{E}^{f_t}[q^2] = \frac{\beta e^{-2\frac{\omega^2}{\gamma}t} + (1 - e^{-2\frac{\omega^2}{\gamma}t})\beta_0}{m\beta_0 \beta \omega^2}. \label{eq:generator_target}
\end{equation}
To apply the formalism of Section~\ref{sec:stochper} to the overdamped dynamics expressed by Eq.~\eqref{eq:langevin}, we first rewrite the latter in the form:
\begin{equation}
\dot{q}(t) = -\frac{\omega^2}{\gamma} q(t) + \sigma \xi(t), \quad q(0) = q_0,
\label{eq:K-L_noise0}
\end{equation}
where $\sigma = \sqrt{2D}$ and with the initial state $q_0$ sampled from the distribution $f_0$ given in Eq.~\eqref{eq:initial_pdf}.\footnote{We note that, within the generator-based formalism of Section~\ref{sec:gener}, the distribution $f_0$ is invariant for the unperturbed stochastic dynamics. This is not the case, however, in the framework of the K--L expansion of Section~\ref{sec:stochper}, where the unperturbed dynamics is deterministic and dissipative and is therefore endowed with a singular invariant measure. In this case, $f_0$ simply represents an arbitrary initial distribution.} The physical model corresponds to an overdamped harmonic oscillator subject, from time $t=0$ onwards, to a white-noise perturbation with assigned strength $D>0$.
Following the procedure outlined in Sec.~\ref{sec:stochper}, we make the dynamical system autonomous by introducing two auxiliary variables, $\theta$ and $\phi$, obeying the harmonic oscillator equations~\eqref{oscillator}, which we recall here:
\begin{equation*}
\begin{cases}
\dot{\theta} = b \phi, \\
\dot{\phi} = -b \theta,
\end{cases}
\end{equation*}
with $b=\pi/(2T)$.
The autonomous system in the augmented phase space, with state vector $\tilde{\Gamma} = (q, \theta, \phi)^\top$, then takes the form
\begin{equation}
\dot{\tilde{\Gamma}} = \begin{pmatrix} 
\dot{q} \\ \dot{\theta} \\ \dot{\phi} 
\end{pmatrix} = \begin{pmatrix}
-\frac{\omega^2}{\gamma} q + \sigma \xi(\theta) \\
b\phi \\
-b\theta
\end{pmatrix}.
\end{equation}

We parametrize the initial conditions of the auxiliary subsystem on the unit circle by a single phase $\psi \sim \mathcal{U}[0, 2\pi)$, setting $\theta_0 = \cos\psi$ and $\phi_0 = -\sin\psi$; this uniform phase distribution corresponds to the density $g_0$ introduced in Sec.~\eqref{sec:stochper}. Integrating the auxiliary subsystem yields the explicit trajectories $\theta(t) = \cos(bt + \psi)$ and $\phi(t) = -\sin(bt + \psi)$.
The noise is expressed via the K--L expansion 
\begin{equation}
\xi(t) = \sqrt{\frac{2}{T}} \sum_{k=1}^{\infty} Z_k \cos(b_k t + \psi_k) = \sqrt{\frac{2}{T}} \sum_{k=1}^{\infty} Z_k\, T_{2k-1}(\theta(t)),
\label{eq:xi_psi}
\end{equation}
where $b_k = (2k - 1)b$, $\psi_k = (2k - 1)\psi$, $T_m$ denotes the $m$-th Chebyshev polynomial of the first kind, so that $\cos(b_k t + \psi_k) = T_{2k-1}(\theta(t))$, and the amplitudes $Z_k \sim \mathcal{N}(0, 1)$ are independent standard normal variables, independent of $q_0$ and $\psi$.\footnote{The expansion~\eqref{eq:xi_psi} is readily obtained from Eq.~\eqref{expansion}, by noticing that, since $\xi$ is real-valued, the coefficients satisfy $\alpha_{-n} = \bar{\alpha}_n$, with $\beta_{-n} = -\beta_n$, so that the terms of opposite index combine into cosines. Setting $\beta_{\pm k} = \pm b_k$ and $\alpha_{\pm k} = Z_k\, e^{\pm i\psi_k}/\sqrt{2T}$, $k \ge 1$, with no $n = 0$ term, the expansion takes the real form~\eqref{eq:xi_psi}.}

Denoting by $\mathbb{E}_Z[\cdot]$ the expectation over the amplitudes alone, one has $\mathbb{E}_Z[Z_k] = 0$ and $\mathbb{E}_Z[Z_j Z_k] = \delta_{jk}$.

Since the frequency and the phase of each mode are the same odd multiple of $b$ and $\psi$, respectively, each mode reads $\cos\big((2k-1)(bt+\psi)\big)$, and the second equality follows from the defining property $T_n(\cos\alpha)=\cos(n\alpha)$ of the Chebyshev polynomials of the first kind~\cite{mason2002chebyshev}. The noise can thus be written in terms of the coordinate $\theta$ only. We denote its initial value by $\xi_0 := \xi(0)$. Details on the K--L expansion are presented in Appendix~\ref{app:appB} \footnote{For $\psi=0$, Eq.~\eqref{eq:xi_psi} reduces to the standard K--L expansion of white noise on $[0,T]$; for a generic phase, the white-noise correlation structure is recovered upon averaging over $\psi$, as is always the case in the expectations considered below.}.

To compute the response formula, we first solve explicitly the first-order linear non-homogeneous ODE for the variable $q$, obtaining
\begin{equation}
q(s) = q_0 e^{-\frac{\omega^2}{\gamma}s} + \sigma \int_{0}^{s} e^{-\frac{\omega^2}{\gamma}(s-\tau)}\xi(\tau)\,\mathrm{d}\tau,
\label{solq}
\end{equation}
where $q(s)$ denotes the $q$-component of the perturbed trajectory $\tilde{S}^s(q_0,\theta_0,\phi_0)$ in the augmented phase space. Regarding $q^2$ as an observable on $\tilde{\mathcal{M}}$, one thus has $q^2(s) = (q^2\circ\tilde{S}^s)(q_0,\theta_0,\phi_0)$; through the noise, $q(s)$ depends on $q_0$, on the initial phase $\psi$ and on the amplitudes $\{Z_k\}_{k\ge1}$. 
Squaring yields\footnote{When the series~\eqref{eq:xi_psi} is truncated to its first $N$ modes, the noise is a smooth function of time for every realization of the amplitudes $\{Z_k\}_{k=1}^{N}$ and of the phase $\psi$, so that Eqs.~\eqref{solq} and~\eqref{eq:squared_trajectory} hold pathwise as ordinary identities. For the full series, $\xi$ is no longer a function; nevertheless, the integral in Eq.~\eqref{solq}, defined as the mean-square limit of its truncated counterparts, exists (see Appendix~\ref{app:appC}), so that $q(s)$ and $q^2(s)$ are well-defined random variables. What remains formal is the initial noise $\xi_0$ entering the dissipation function~\eqref{omega_definition}: products involving $\xi_0$ acquire a precise meaning only after averaging over the amplitudes and the phase, through the noise autocorrelation $C_T$, with the K--L series integrated term by term.}

\begin{align}
q^2(s) &= q_0^2 e^{-2\frac{\omega^2}{\gamma}s} 
+ 2\sigma q_0 e^{-\frac{\omega^2}{\gamma}s} \int_0^s e^{-\frac{\omega^2}{\gamma}(s-\tau)} \xi(\tau) \, \mathrm{d}\tau \nonumber\\
&+ \sigma^2 \int_0^s \!\! \int_0^s e^{-\frac{\omega^2}{\gamma}(2s - u_1 - u_2)} \xi(u_1)\xi(u_2) \, \mathrm{d}u_1 \, \mathrm{d}u_2.
\label{eq:squared_trajectory}
\end{align}
Next, we define the probability density $\tilde{f}_0(q_0,\theta_0,\phi_0) = f_0(q_0)\,g_0(\theta_0,\phi_0)$, where
\begin{equation}
g_0(\theta_0, \phi_0) = \frac{1}{2\pi}\, \delta\Big(\sqrt{\theta_0^2 + \phi_0^2} - 1\Big)
\label{eq:g0}
\end{equation}
is the uniform density on the unit circle. As in Section~\ref{sec:stochper}, $\hat{\mathbb{E}}^{f}[\cdot]$ denotes the expectation with respect to a density $f$ for a fixed realization of the amplitudes $\{Z_k\}_{k\ge1}$, and $\mathbb{E}^{f}[\cdot]$ the double expectation over the amplitudes and with respect to $f$. For densities that do not involve the noise, such as $f_0$, the two coincide.

Under the assumption that $\tilde{f}_0$ evolves according to Liouville's equation, the dissipation function $\tilde{\Omega}^{\tilde{f}_0} = -\nabla_{\tilde\Gamma} \cdot \dot{\tilde\Gamma} - \dot{\tilde\Gamma} \cdot \nabla_{\tilde\Gamma} \ln \tilde{f}_0$ reads
\begin{equation}
\tilde{\Omega}^{\tilde{f}_0}(q_0,\theta_0,\phi_0) = \frac{\omega^2}{\gamma} - \frac{\omega^4}{\gamma}m\beta_0 q_0^2 + \omega^2 m\beta_0 \sigma\, q_0\, \xi_0(\theta_0).
\label{omega_definition}
\end{equation}
The exact response formula for the observable $\mathcal{O}=q^2$ is therefore obtained from Eq.~\eqref{eq:ERTst} in the form
\begin{equation}
\mathbb{E}^{\tilde{f}_t} [q^2] = \mathbb{E}^{f_0}[q^2] + \int_{0}^{t} \mathbb{E}^{\tilde{f}_0}\!\left[\tilde{\Omega}^{\tilde{f}_0}\, q^2(s) \right] \dd s.
\label{response_formula}
\end{equation}
Inserting the trajectory-based expression~\eqref{eq:squared_trajectory} into Eq.~\eqref{response_formula} and exploiting the independence of $q_0$ from the noise together with the Gaussian moments of $f_0$ yields
\begin{equation}
\mathbb{E}^{\tilde{f}_0}\!\left[\tilde{\Omega}^{\tilde{f}_0}\, q^2(s)\right] = -\frac{2}{\gamma m \beta_0} e^{-2\frac{\omega^2}{\gamma}s} + 2 \sigma^2 e^{-\frac{\omega^2}{\gamma}s} \int_{0}^{s} e^{-\frac{\omega^2}{\gamma}(s-\tau)}\, C_T(\tau) \, \dd\tau,
\label{eq:cross_term_integral}
\end{equation}
where
\begin{equation}
C_T(\tau) := \mathbb{E}^{g_0}[\xi(\tau)\,\xi_0]
\label{eq:CT}
\end{equation}
is the noise autocorrelation, averaged over both the initial phase and the amplitudes. Details on the derivation of Eq.~\eqref{eq:cross_term_integral} are given in Appendix~\ref{app:appC}.

\begin{figure}[h]
\centering
\includegraphics[width=0.49\linewidth]{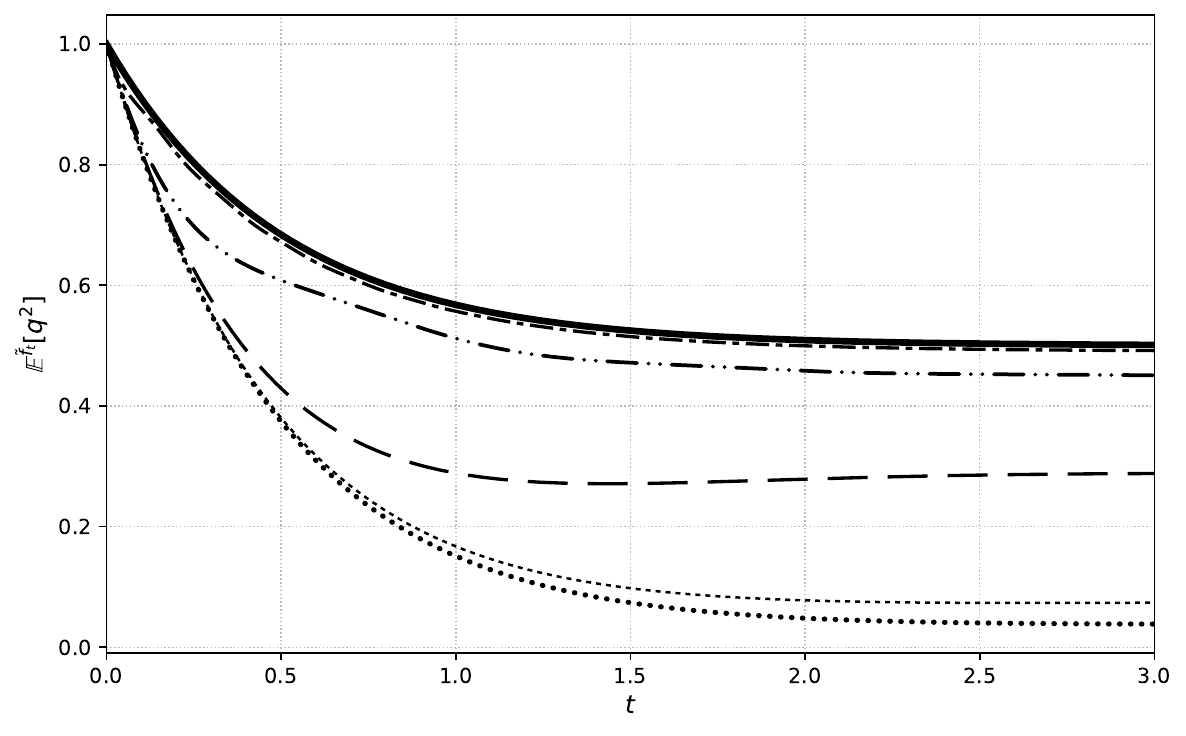}
\hfill
\includegraphics[width=0.49\linewidth]{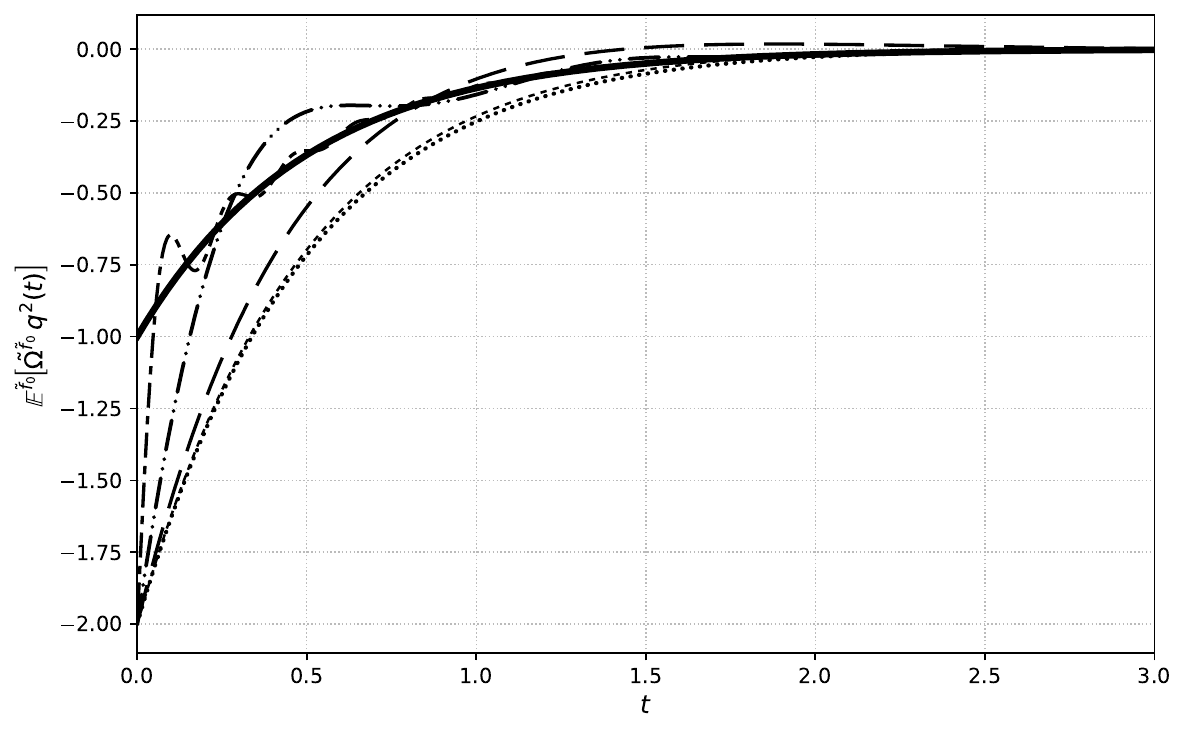}
\caption{\textit{Left panel}: Time evolution of $\mathbb{E}^{\tilde{f}_t}[q^2]^{(N)}$ for $\beta = 2 \beta_0$, $\omega=\gamma$ and $T=25$. Lines represent the analytical result~\eqref{eq:q2_N} for various truncation orders: $N = 1$ (dotted), $N = 2$ (short-dashed), $N = 10$ (long-dashed), $N = 50$ (long dash followed by two dots) and $N = 270$ (long dash followed by a single dot). The solid black line indicates the generator-based result~\eqref{eq:generator_target}. \textit{Right panel}: Time evolution of the correlation function $\mathbb{E}^{\tilde{f}_0}\big[\tilde{\Omega}^{\tilde{f}_0}\,q^2(t)\big]^{(N)}$, Eq.~\eqref{eq:corr_N}, for the same truncation orders. Parameter values and line styles are as in the left panel.}
\label{fig:K-L_FP_convergence}
\end{figure}

\begin{figure}[htbp]
\centering
\includegraphics[width=0.55\linewidth]{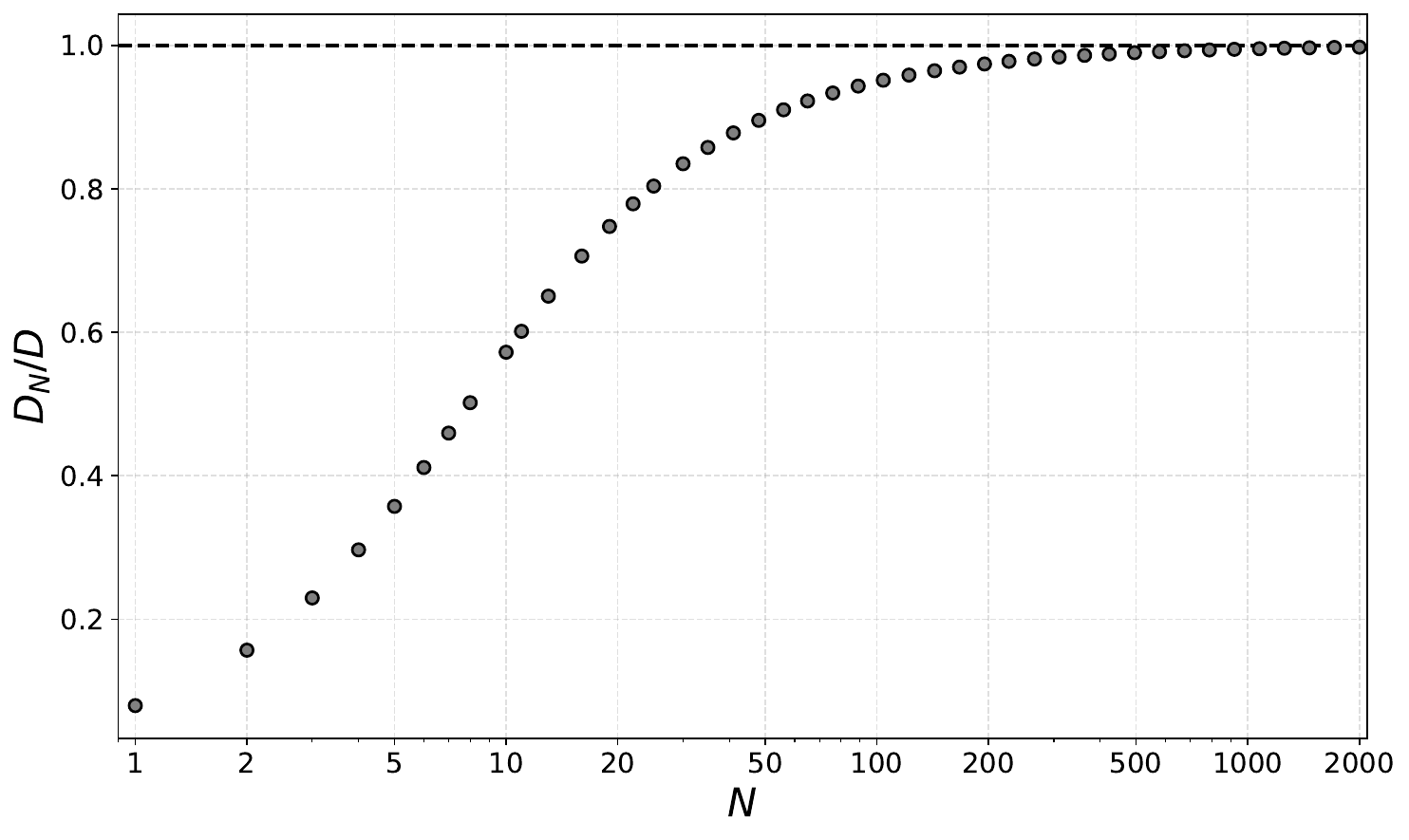}
\caption{Behavior of the ratio of the effective noise amplitude $D_N$, defined in Eq.~\eqref{eq:DN}, to the theoretical value $D = (m\beta \gamma)^{-1}$, plotted as a function of the number of K--L modes $N$. 
The parameters are set to the same values as in Fig.~\ref{fig:K-L_FP_convergence}.}
\label{fig:ratio}
\end{figure}

\subsection{Finite order truncations}
\label{sec:trunc}

In view of numerical implementations, it is often convenient to consider finite-order truncations of expansion~\eqref{eq:xi_psi}. 
We denote by $\xi^{(N)}$ the truncation of the series~\eqref{eq:xi_psi} to its first $N$ modes, written as follows:
\begin{equation}
\xi^{(N)}(t) = \sqrt{\frac{2}{T}} \sum_{k=1}^{N} Z_k \cos(b_k t + \psi_k) = \sqrt{\frac{2}{T}} \sum_{k=1}^{N} Z_k T_{2k-1}(\theta(t)), \qquad N \ge 1,
\label{eq:xi_psi_N}
\end{equation}
see Eq.~\eqref{eq:xi_psi} for the definition of the symbols, and by $\xi_0^{(N)} := \xi^{(N)}(0)$ its initial value.
Accordingly, the noise autocorrelation~\eqref{eq:CT} is replaced by
\begin{equation}
C_T^{(N)}(\tau) := \mathbb{E}^{g_0}\big[\xi^{(N)}(\tau)\,\xi^{(N)}_0\big] = \frac{1}{T}\sum_{k=1}^{N}\cos(b_k\tau),
\label{eq:CT_N}
\end{equation}
as derived in Appendix~\ref{app:appD}. We also denote by $\mathbb{E}^{\tilde f}[\cdot]^{(N)}$ the double expectation introduced in Section~\ref{sec:stochper}, evaluated with $\xi^{(N)}$ in place of $\xi$, both in the trajectory $q(s)$ and in the dissipation function~\eqref{omega_definition}. Inserting Eq.~\eqref{eq:CT_N} into Eq.~\eqref{eq:cross_term_integral}, and the result into Eq.~\eqref{response_formula}, yields an exact analytical formula for arbitrary $N \ge 1$:
\begin{equation}
\mathbb{E}^{\tilde{f}_t}[q^2]^{(N)} = \frac{e^{-2\frac{\omega^2}{\gamma}t}}{m\beta_0\omega^2} + \frac{2\gamma^2\sigma^2}{T}\, e^{-\frac{\omega^2}{\gamma}t} \sum_{k=1}^{N} \frac{\cosh\!\left(\frac{\omega^2}{\gamma}t\right) - \cos(b_k t)}{\omega^4 + \gamma^2 b_k^2}.
\label{eq:q2_N}
\end{equation}
For $N=1$ and $N=2$, formula~\eqref{eq:q2_N} reads explicitly
\begin{align}
\mathbb{E}^{\tilde{f}_t}[q^2]^{(1)} &= \frac{e^{-2\frac{\omega^2}{\gamma}t}}{m\beta_0\omega^2} + \frac{2\gamma^2\sigma^2}{T}\, e^{-\frac{\omega^2}{\gamma}t}\, \frac{\cosh\!\left(\frac{\omega^2}{\gamma}t\right) - \cos(bt)}{\omega^4 + \gamma^2 b^2},
\label{eq:q2_N1} \\
\mathbb{E}^{\tilde{f}_t}[q^2]^{(2)} &= \mathbb{E}^{\tilde{f}_t}[q^2]^{(1)} + \frac{2\gamma^2\sigma^2}{T}\, e^{-\frac{\omega^2}{\gamma}t}\, \frac{\cosh\!\left(\frac{\omega^2}{\gamma}t\right) - \cos(3bt)}{\omega^4 + 9\gamma^2 b^2}.
\label{eq:q2_N2}
\end{align}
Correspondingly, the correlation function shown in the right panel of Fig.~\ref{fig:K-L_FP_convergence}, obtained by inserting Eq.~\eqref{eq:CT_N} into Eq.~\eqref{eq:cross_term_integral}, admits the closed form:
\begin{equation}
\mathbb{E}^{\tilde{f}_0}\!\left[\tilde{\Omega}^{\tilde{f}_0} q^2(t)\right]^{(N)}
= -\frac{2}{\gamma m \beta_0}\, e^{-2\frac{\omega^2}{\gamma}t}
+ \frac{2\gamma\sigma^2}{T}\, e^{-\frac{\omega^2}{\gamma}t}
\sum_{k=1}^{N} \frac{\omega^2\left[\cos(b_k t) - e^{-\frac{\omega^2}{\gamma}t}\right] + \gamma b_k \sin(b_k t)}{\omega^4 + \gamma^2 b_k^2}.
\label{eq:corr_N}
\end{equation}
For $N=1$ and $N=2$, the correlation function~\eqref{eq:corr_N} reads explicitly:
\begin{align}
\mathbb{E}^{\tilde{f}_0}\!\left[\tilde{\Omega}^{\tilde{f}_0} \,q^2(t)\right]^{(1)} &= -\frac{2}{\gamma m \beta_0}\, e^{-2\frac{\omega^2}{\gamma}t} + \frac{2\gamma\sigma^2}{T}\, e^{-\frac{\omega^2}{\gamma}t}\, \frac{\omega^2\left[\cos(bt) - e^{-\frac{\omega^2}{\gamma}t}\right] + \gamma b \sin(bt)}{\omega^4 + \gamma^2 b^2},
\label{eq:corr_N1} \\
\mathbb{E}^{\tilde{f}_0}\!\left[\tilde{\Omega}^{\tilde{f}_0} \,q^2(t)\right]^{(2)} &= \mathbb{E}^{\tilde{f}_0}\!\left[\tilde{\Omega}^{\tilde{f}_0} \,q^2(t)\right]^{(1)} + \frac{2\gamma\sigma^2}{T}\, e^{-\frac{\omega^2}{\gamma}t}\, \frac{\omega^2\left[\cos(3bt) - e^{-\frac{\omega^2}{\gamma}t}\right] + 3\gamma b \sin(3bt)}{\omega^4 + 9\gamma^2 b^2}.
\label{eq:corr_N2}
\end{align}


The left panel of Fig.~\ref{fig:K-L_FP_convergence} shows $\mathbb{E}^{\tilde{f}_t}[q^2]^{(N)}$ for different values of the truncation order $N$, together with the generator-based expectation $\mathbb{E}^{f_t}[q^2]$, which serves as a reference. As more K--L modes are added to the 
expansion, we observe that the truncated analytic expectation functions progressively approaches the generator-based result~\eqref{eq:generator_target}; convergence of the full K--L expansion to the latter will be proved in Section~\ref{sec:limit}, in the limit $T\to\infty$.
The behavior of the correlation functions $\mathbb{E}^{\tilde{f}_0}\!\left[\tilde{\Omega}^{\tilde{f}_0}\,q^2(t)\right]^{(N)}$ is displayed in the right panel of the same figure. 

Convergence toward the generator-based result can also be monitored through the behavior, for growing values of $N$, of the quantity
\begin{equation}
D_N := \frac{\omega^2}{\gamma} \lim_{t \to \infty} \mathbb{E}^{\tilde{f}_t}[q^2]^{(N)},
\label{eq:DN}
\end{equation}
which corresponds to the \emph{effective} diffusion coefficient associated with the $N$-th order truncation of the K--L expansion, recalling that, for a noise term, $\lim_{t \to \infty} \mathbb{E}[q^2]$ has also to be taken.
Since the sum in Eq.~\eqref{eq:q2_N} is finite, the limit $t\to\infty$ can be taken term by term, each term $e^{-\frac{\omega^2}{\gamma}t}\big[\cosh(\frac{\omega^2}{\gamma}t)-\cos(b_k t)\big]$ tending to $1/2$. Using $\sigma^2 = 2D = 2(m\beta\gamma)^{-1}$, Eq.~\eqref{eq:DN} then yields, for arbitrary $N \ge 1$,
\begin{equation}
D_N = \frac{2\omega^2}{m\beta T} \sum_{k=1}^{N} \frac{1}{\omega^4 + \gamma^2 b_k^2},
\qquad \text{i.e.} \qquad
\frac{D_N}{D} = \frac{2\gamma\omega^2}{T} \sum_{k=1}^{N} \frac{1}{\omega^4 + \gamma^2 b_k^2}.
\label{eq:DN_explicit}
\end{equation}
Each K--L mode thus provides an independent, additive contribution to the effective diffusion coefficient, whose weight $(\omega^4+\gamma^2 b_k^2)^{-1}$ decreases with the mode frequency $b_k$: modes with $b_k \gg \omega^2/\gamma$, i.e.\ much faster than the relaxation of the oscillator, contribute only as $b_k^{-2}$. In the limit $N\to\infty$, the series can be summed in closed form,\footnote{This follows from the identity $\sum_{k\ge1}\big[c^2+(2k-1)^2\big]^{-1} = \frac{\pi}{4c}\tanh\frac{\pi c}{2}$~\cite{gradshteyn2007table}, with $c = \omega^2/(\gamma b)$ and $b = \pi/(2T)$.} yielding
\begin{equation}
\lim_{N\to\infty} D_N = D \tanh\!\left(\frac{\omega^2}{\gamma}\,T\right).
\label{eq:Dinf}
\end{equation}
The behavior of $D_N$ as a function of $N$ is shown in Figure~\ref{fig:ratio}. For fixed $T$, Eq.~\eqref{eq:Dinf} shows that the diffusion coefficient $D$ is recovered up to an exponentially small correction, $1-\tanh(\omega^2 T/\gamma) \simeq 2e^{-2\omega^2 T/\gamma}$, which is negligible for the parameters considered here, and exactly in the limit $T\to\infty$. 
Formula~ \eqref{eq:DN_explicit} also highlights that the order of the limits is essential. At fixed $N$, all frequencies $b_k = (2k-1)\pi/(2T)$ vanish as $T\to\infty$, and Eq.~\eqref{eq:DN_explicit} gives $D_N/D \simeq 2\gamma N/(\omega^2 T) \to 0$: a finite number of modes cannot represent white noise over an arbitrarily long time window. The white-noise limit therefore requires $N\to\infty$ to be taken before $T\to\infty$, in agreement with the general remark of Section~\ref{sec:stochper} that these limits do not necessarily commute. 

We investigate the large-$T$ limit for the expectation $\mathbb{E}^{\tilde{f}_t}[q^2]$ in Section~\ref{sec:limit} below.

\subsection{Large $T$ limit}
\label{sec:limit}

We finally show  explicitly that the response formula~\eqref{response_formula}, which includes the full K--L expansion of the noise,  recovers the generator-based expression~\eqref{eq:generator_target} in the limit $T\to\infty$. This provides support for the applicability of ERT to more complex as well as to non-markovian processes.

Heuristically, the result follows from the behavior of the noise autocorrelation. Taking $N\to\infty$ in Eq.~\eqref{eq:CT_N} and introducing the spacing $\Delta b := b_{k+1}-b_k = \pi/T$ between consecutive mode frequencies, the autocorrelation~\eqref{eq:CT} can be written as a Riemann sum,
\begin{equation}
C_T(\tau) = \frac{1}{T}\sum_{k=1}^{\infty}\cos(b_k\tau) = \frac{1}{\pi}\sum_{k=1}^{\infty}\cos(b_k\tau)\,\Delta b.
\label{eq:CT_riemann}
\end{equation}
As $T\to\infty$, the spacing $\Delta b$ vanishes and, as detailed in Appendix~\ref{app:appD}, the sum formally converges to
\begin{equation}
\lim_{T\to\infty}C_T(\tau) = \frac{1}{\pi}\int_0^\infty\cos(\nu\tau)\,\mathrm{d}\nu = \delta(\tau),
\label{eq:dirac}
\end{equation}
where $\delta(\cdot)$ denotes the Dirac delta, as is characteristic of white noise~\cite{risken1996fokker}. Inserting Eq.~\eqref{eq:dirac} into Eq.~\eqref{eq:cross_term_integral}, with weight $1/2$ for the delta located at the lower boundary of the integration domain, and then into Eq.~\eqref{response_formula}, yields Eq.~\eqref{eq:generator_target} after straightforward integration.

However, this argument is only formal because $C_T$ is not a function, hence the limit~\eqref{eq:dirac} holds only in the distributional sense. A rigorous argument comes with the following proposition, in which the integral of $C_T$ against a kernel is understood term by term through the series~\eqref{eq:CT_riemann}.

\begin{proposition}
\label{prop:prop1}
Let the integral of $C_T$ in Eq.~\eqref{eq:cross_term_integral} be defined term by term through the series~\eqref{eq:CT_riemann}. Then, for every $t \ge 0$,
\begin{equation}
\lim_{T\to\infty}\mathbb{E}^{\tilde{f}_t}[q^2] = \frac{\beta e^{-2\frac{\omega^2}{\gamma}t} + \left(1 - e^{-2\frac{\omega^2}{\gamma}t}\right)\beta_0}{m\beta_0 \beta \omega^2}
\label{eq:KL_goal}
\end{equation}
\end{proposition}

\begin{proof}
Let $a = \omega^2/\gamma$, and recall that $\Delta b = \pi/T$, so that $b_k = (2k-1)\pi/(2T)$ is the midpoint of the interval $I_k = [(k-1)\Delta b, k\Delta b]$; the intervals $I_k$, $k \ge 1$, form a partition of the half-line $[0,\infty)$.
Inserting Eq.~\eqref{eq:cross_term_integral} into Eq.~\eqref{response_formula}, $\mathbb{E}^{\tilde{f}_t}[q^2]$ splits into a noise-independent and a noise-dependent part. The former, obtained by integrating the first term of Eq.~\eqref{eq:cross_term_integral} over $s\in[0,t]$ and adding $\mathbb{E}^{f_0}[q^2] = (m\beta_0\omega^2)^{-1}$, gives
\begin{equation*}
\frac{1}{m\beta_0\omega^2} - \frac{2}{\gamma m\beta_0}\int_0^t e^{-2as}\,\mathrm{d}s = \frac{e^{-2at}}{m\beta_0\omega^2},
\end{equation*}
independently of $T$. The latter reads
\begin{equation*}
2\sigma^2\int_0^t e^{-as}\int_0^s e^{-a(s-\tau)}\,C_T(\tau)\,\mathrm{d}\tau\,\mathrm{d}s,
\end{equation*}
where the double integral runs over the triangle $0 \le \tau \le s \le t$, since $\tau \in [0,s]$ and $s \in [0,t]$. Exchanging the order of the time integrations, the noise-dependent part equals
\begin{equation*}
2\sigma^2\int_0^t K_t(\tau)\,C_T(\tau)\,\mathrm{d}\tau,
\qquad
K_t(\tau) := e^{a\tau}\int_\tau^t e^{-2as}\,\mathrm{d}s = \frac{e^{-a\tau} - e^{-a(2t-\tau)}}{2a}.
\end{equation*}

According to the definition adopted in the statement of the Proposition, the integral of $C_T$ against $K_t$ is obtained by integrating the series~\eqref{eq:CT_riemann} term by term. Since $1/T = \Delta b/\pi$,
\begin{equation*}
\int_0^t K_t(\tau)\,C_T(\tau)\,\mathrm{d}\tau = \frac{1}{T}\sum_{k=1}^{\infty}\int_0^t K_t(\tau)\cos(b_k\tau)\,\mathrm{d}\tau = \frac{1}{\pi}\sum_{k=1}^{\infty} F_t(b_k)\,\Delta b,
\end{equation*}
where $F_t(\nu) := \int_0^t K_t(\tau)\cos(\nu\tau)\,\mathrm{d}\tau$, $\nu \ge 0$, is the cosine transform of the kernel. Evaluating the elementary integrals $\int_0^t e^{\mp a\tau}\cos(\nu\tau)\,\mathrm{d}\tau$ and combining them, the terms proportional to $\sin(\nu t)$ cancel, and one finds
\begin{equation}
F_t(\nu) = 
\frac{e^{-at}\big(\cosh(at) - \cos(\nu t)\big)}{a^2 + \nu^2}.
\label{Ft}
\end{equation}
Since $e^{-at}\cosh(at) = (1+e^{-2at})/2 \le 1$ and $e^{-at}|\cos(\nu t)| \le 1$, one has
\begin{equation*}
0 \le F_t(\nu) \le \frac{2}{a^2+\nu^2}.
\end{equation*}
Since $b_k$ grows linearly in $k$, the terms of the series are therefore $O(k^{-2})$, and the series converges absolutely. The same bound also shows that $\int_0^\infty F_t(\nu)\,\mathrm{d}\nu$ is finite. Moreover, $F_t$ is smooth, and its derivative,
\begin{equation*}
F_t'(\nu) = \frac{t\,e^{-at}\sin(\nu t)}{a^2+\nu^2} - \frac{2\nu\,e^{-at}\big(\cosh(at)-\cos(\nu t)\big)}{(a^2+\nu^2)^2},
\end{equation*}
satisfies $|F_t'(\nu)| \le (t + 2/a)/(a^2 + \nu^2)$, as follows from the previous bound together with $a^2 + \nu^2 \ge 2a\nu$, which gives $4\nu/(a^2+\nu^2)^2 \le (2/a)/(a^2+\nu^2)$.

Next, we show that the Riemann sum $\sum_k F_t(b_k)\,\Delta b$ converges to $\int_0^\infty F_t(\nu)\,\mathrm{d}\nu$ as $T\to\infty$. Splitting the integral over the intervals $I_k$, one has $\int_0^\infty F_t(\nu)\,\mathrm{d}\nu = \sum_{k\ge1}\int_{I_k}F_t(\nu)\,\mathrm{d}\nu$. Since this series and the Riemann sum $\sum_{k\ge1}F_t(b_k)\,\Delta b$ both converge, their difference can be computed interval by interval:
\begin{equation*}
\sum_{k=1}^{\infty} F_t(b_k)\,\Delta b - \int_0^\infty F_t(\nu)\,\mathrm{d}\nu = \sum_{k=1}^{\infty}\left(F_t(b_k)\,\Delta b - \int_{I_k} F_t(\nu)\,\mathrm{d}\nu\right) = \sum_{k=1}^{\infty}\int_{I_k}\big(F_t(b_k) - F_t(\nu)\big)\,\mathrm{d}\nu,
\end{equation*}
where we used that $I_k$ has length $\Delta b$, so that $F_t(b_k)\,\Delta b = \int_{I_k}F_t(b_k)\,\mathrm{d}\nu$. For any $\nu \in I_k$, the fundamental theorem of calculus gives
\begin{equation*}
\big|F_t(b_k) - F_t(\nu)\big| = \left|\int_\nu^{b_k} F_t'(\mu)\,\mathrm{d}\mu\right| \le \int_{I_k}|F_t'(\mu)|\,\mathrm{d}\mu,
\end{equation*}
where the bound does not depend on $\nu$. Integrating it over $I_k$, the error on each interval satisfies
\begin{equation*}
\left|\int_{I_k}\big(F_t(b_k) - F_t(\nu)\big)\,\mathrm{d}\nu\right| \le \Delta b\int_{I_k}|F_t'(\mu)|\,\mathrm{d}\mu.
\end{equation*}
Summing over $k$, the triangle inequality leads to
\begin{equation*}
\left|\sum_{k=1}^{\infty} F_t(b_k)\,\Delta b - \int_0^\infty F_t(\nu)\,\mathrm{d}\nu\right| \le \Delta b\sum_{k=1}^{\infty}\int_{I_k}|F_t'(\mu)|\,\mathrm{d}\mu = \frac{\pi}{T}\int_0^\infty |F_t'(\nu)|\,\mathrm{d}\nu.
\end{equation*}
Finally, using the bound on $F_t'$ one obtains
\begin{equation}
\left|\sum_{k=1}^{\infty} F_t(b_k)\,\Delta b - \int_0^\infty F_t(\nu)\,\mathrm{d}\nu\right| \le \frac{\pi^2}{2aT}\left(t + \frac{2}{a}\right) \xrightarrow[T\to\infty]{} 0.
\label{bound}
\end{equation}

It remains to evaluate the limiting integral. Using $\int_0^\infty (a^2+\nu^2)^{-1}\,\mathrm{d}\nu = \pi/(2a)$ and $\int_0^\infty \cos(\nu t)\,(a^2+\nu^2)^{-1}\,\mathrm{d}\nu = \pi e^{-at}/(2a)$ for $t \ge 0$, integration of $F_t$ in Eq.~\eqref{Ft} gives
\begin{equation*}
\int_0^\infty F_t(\nu)\,\mathrm{d}\nu = \frac{\pi}{2a}\,e^{-at}\big(\cosh(at) - e^{-at}\big) = \frac{\pi}{4a}\big(1 - e^{-2at}\big).
\end{equation*}
Hence the noise-dependent part converges to $2\sigma^2(1 - e^{-2at})/(4a) = (1 - e^{-2at})/(m\beta\omega^2)$, where $\sigma^2 = 2(m\beta\gamma)^{-1}$ was used. Adding the noise-independent part $e^{-2at}/(m\beta_0\omega^2)$ yields the claim.
\end{proof}

Some technical remarks are in order. We observe, first, that the proof makes the heuristic argument discussed in Appendix~\ref{app:appD} precise: indeed the limit $T\rightarrow \infty$ of the Riemann sum is taken not on $C_T$ itself (where it is only formal, see Eq.~\eqref{formal}), but on $F_t$, i.e. the cosine transform of the kernel $K_t$ against which $C_T$ is integrated. Moreover, the identity $\pi^{-1}\int_0^\infty F_t(\nu)\,\mathrm{d}\nu = \frac{1}{2}K_t(0)$ reproduces the weight $1/2$ of the Dirac delta located at the boundary of the integration domain, which is used in the heuristic derivation discussed in Appendix~\ref{app:appD}.

\vskip 5pt
\noindent

\noindent

\noindent
{\bf Remark.} {\em
Eq.~\eqref{eq:KL_goal} coincides with the generator-based result~\eqref{eq:generator_target}, i.e.\ $\lim_{T\to\infty}\mathbb{E}^{\tilde{f}_t}[q^2] = \mathbb{E}^{f_t}[q^2]$, although the two expectations are taken with respect to different densities: $\tilde{f}_t$, defined on the augmented phase space, and $f_t$, defined on the state space. The exact response is thus recovered via two conceptually distinct routes: the generator-based approach of Section~\ref{sec:gener} 
operates within the state space of a Markovian stochastic process, whereas the ERT framework formulates the response within the augmented phase space of Section~\ref{sec:stochper}, 
in which the dynamical equations ara autonomous. This encompasses also single realizations of the forcing and finite truncations of the KL expansion.}

\section{Comparison with Linear Response}
\label{sec:LRT}

In this section, we compare the predictions of ERT against linear response theory for the overdamped Langevin dynamics introduced in Section~\ref{sec:LOD}. To facilitate a direct analytical comparison between the two approaches, we follow the procedure outlined in \cite{amadori2022exact} (which is applicable, in the linear regime, even to non-Hamiltonian perturbations) and regard $\sigma$ as the perturbation parameter. While linear response assumes a weak perturbation regime, ERT holds without restriction on the magnitude of $\sigma$.
The derivation of the linear response formula as the first-order expansion of the exact response is presented in Appendix~\ref{app:appE}.

A fundamental distinction between linear and exact response formalisms lies in the
dynamics that is used: while linear response evaluates correlation functions along the unperturbed flow $\tilde{S}_0^t$ (see Eq.~\eqref{responselin}), ERT propagates observable values via the full perturbed flow $\tilde{S}^t$. Consequently, even when starting from an identical initial distribution $\tilde{f}_0$, the probability densities evolve differently in the two
frameworks.
To distinguish them, we denote by $\bar{f}_t$ the density evolved under the linear response approximation at time $t$, reserving $\tilde{f}_t$ for the exact density at time $t$ driven by the full dynamics.

In the traditional use of 
linear response equations
\eqref{linresponse}--\eqref{responselin}, the initial density is invariant under the unperturbed dynamics (classical ensembles and non-driven Hamiltonian dynamics), which guarantees that the dissipation function is indeed proportional to the external field and to the dissipative flux. For generic dynamical systems, that do not need to represent transport of particles,
$f_0$ does not need to be invariant under the unperturbed 
flow $\tilde{S}^s_0$. 
Therefore, we are not committed to an initial distribution that is invariant for the unperturbed dynamics, and
we take the linear response formula as the first-order expansion of the exact response in $\sigma$, using the perturbed flow. As shown in Appendix~\ref{app:appE}, when $f_0$ is not invariant this expansion contains, besides the usual contributions, a term accounting for the unperturbed evolution of $f_0$ plus a term linear in the perturbation, that accounts for the deviation of the trajectory from its unperturbed realization 
(see Eq.~\eqref{eq:app_ERT_first_order} therein). In the present model, however, the latter vanishes for every observable depending on $q$ only. Consequently, the linear response prediction for $\mathcal{O} = q^2$ takes the classical form, in which the dissipation function is correlated with the observable propagated along the unperturbed flow, $q^2 \circ \tilde{S}_0^s = q_0^2 e^{-2\frac{\omega^2}{\gamma}s}$:
\begin{equation}
\mathbb{E}^{\bar{f}_t}[q^2] = \mathbb{E}^{f_0}[q^2] + \int_0^t \mathbb{E}^{\tilde{f}_0}\!\left[\tilde{\Omega}^{\tilde{f}_0}\left(q^2 \circ \tilde{S}_0^s\right)\right] \mathrm{d}s,
\label{eq:LRT_formula}
\end{equation}
which reduces to Eq.~\eqref{linresponse} when $f_0$ is invariant.
%
Using Eq.~\eqref{omega_definition}, the integrand reads
\begin{align}
\mathbb{E}^{\tilde{f}_0}\!\left[\tilde{\Omega}^{\tilde{f}_0}\left(q^2 \circ \tilde{S}_0^s\right)\right] &= \mathbb{E}^{\tilde{f}_0}\!\left[ \left( \frac{\omega^2}{\gamma} - \frac{\omega^4}{\gamma}m\beta_0 q_0^2 + \omega^2 m\beta_0 q_0 \sigma \xi_0 \right) q_0^2 e^{-2\frac{\omega^2}{\gamma}s} \right] \nonumber \\
&= e^{-2\frac{\omega^2}{\gamma}s} \left( \frac{\omega^2}{\gamma} \mathbb{E}^{f_0}[q_0^2] - \frac{\omega^4}{\gamma}m\beta_0 \mathbb{E}^{f_0}[q_0^4] + \omega^2 m\beta_0 \sigma \mathbb{E}^{f_0}[q_0^3]\mathbb{E}^{g_0}[\xi_0] \right).
\end{align}
and this is the only step to be taken.
Because we chose
$f_0$ in Eq.~\eqref{eq:initial_pdf} as
a centered Gaussian distribution with parameter $\beta_0$, all odd moments vanish ($\mathbb{E}^{f_0}[q_0^3] = 0$), and the even moments are given by $\mathbb{E}^{f_0}[q_0^2] = (m\beta_0\omega^2)^{-1}$ and $\mathbb{E}^{f_0}[q_0^4] = 3(m\beta_0\omega^2)^{-2}$. Furthermore, the initial noise vanishes upon averaging, $\mathbb{E}^{g_0}[\xi_0] = 0$, with the series~\eqref{eq:xi_psi} averaged term by term as in Prop.~\ref{prop:prop1}. Therefore, the term proportional to the perturbation $\sigma$ vanishes entirely, leaving:
\begin{equation}
\mathbb{E}^{\tilde{f}_0}\!\left[\tilde{\Omega}^{\tilde{f}_0}\left(q^2 \circ \tilde{S}_0^s\right)\right] = e^{-2\frac{\omega^2}{\gamma}s} \left( \frac{1}{\gamma m \beta_0} - \frac{3}{\gamma m \beta_0} \right) = - \frac{2}{\gamma m \beta_0} e^{-2\frac{\omega^2}{\gamma}s}.
\end{equation}
Integrating this result over the interval $s \in [0, t]$ gives the response integral:
\begin{align}
\int_{0}^{t} \mathbb{E}^{\tilde{f}_0}\!\left[\tilde{\Omega}^{\tilde{f}_0}\left(q^2 \circ \tilde{S}_0^s\right)\right] \mathrm{d}s &= - \frac{2}{\gamma m \beta_0} \int_{0}^{t} e^{-2\frac{\omega^2}{\gamma}s} \, \mathrm{d}s \nonumber \\
&= \frac{1}{m\beta_0\omega^2} \left( e^{-2\frac{\omega^2}{\gamma}t} - 1 \right).
\end{align}
Finally, substituting this integral back into Eq.~\eqref{eq:LRT_formula} alongside the initial expectation $\mathbb{E}^{f_0}[q^2] = (m\beta_0\omega^2)^{-1}$, we obtain the linear response prediction for the observable $\mathcal{O}=q^2$:
\begin{equation}
\mathbb{E}^{\bar{f}_t}[q^2] = \frac{1}{m\beta_0\omega^2} + \frac{1}{m\beta_0\omega^2} \left( e^{-2\frac{\omega^2}{\gamma}t} - 1 \right) = \frac{1}{m\beta_0\omega^2} e^{-2\frac{\omega^2}{\gamma}t}.
\label{eq:lin_response_result}
\end{equation}
Remarkably, the linear response formalism predicts a purely deterministic exponential decay, entirely missing the diffusion term induced by the noise. This is not specific to $q^2$: as shown in Appendix~\ref{app:appE}, for every observable depending on $q$ only, the first-order noise contributions vanish and the linear response reduces to the unperturbed relaxation of the initial density (see Eq.~\eqref{eq:app_LRT_unperturbed}). Since the noise does not contribute at first order, the linear response prediction~\eqref{eq:lin_response_result} is independent of $T$ (and, as shown below, of the truncation order $N$), and thus requires no large-$T$ limit. Capturing the effect of the noise thus requires going beyond first order in $\sigma$.

Conversely, the exact response contains a noise contribution of order $\sigma^2$, as seen by writing the large-$T$ limit of $\mathbb{E}^{\tilde{f}_t}[q^2]$, which by virtue of Prop.~\ref{prop:prop1} coincides with the generator-based result~\eqref{eq:generator_target}, as
\begin{equation}
\lim_{T\to\infty}\mathbb{E}^{\tilde{f}_t}[q^2] = \mathbb{E}^{f_t}[q^2] = \frac{e^{-2\frac{\omega^2}{\gamma}t}}{m\beta_0 \omega^2} + \frac{\sigma^2 \gamma}{2 \omega^2}\left(1 - e^{-2\frac{\omega^2}{\gamma}t}\right).
\label{eq:generator_target2}
\end{equation}
Since this contribution is quadratic in $\sigma$, it is absent from a response calculation truncated at first order in $\sigma$. The difference between the exact and the linear response predictions thus reads
\begin{equation}
\Delta\mathbb{E}[q^2] := \mathbb{E}^{f_t}[q^2] - \mathbb{E}^{\bar{f}_t}[q^2] = \frac{\gamma \sigma^2}{2 \omega^2} \left( 1 - e^{-2\frac{\omega^2}{\gamma}t} \right),
\label{eq:Delta}
\end{equation}
which approaches the asymptotic value $\gamma D/\omega^2 = (m\beta\omega^2)^{-1}$, 
on the relaxation timescale $\tau_r := \gamma/(2\omega^2)$ of the second moment \cite{balakrishnan2020elements}.

Recalling the definition of $\xi^{(N)}(t)$ in Eq.~\eqref{eq:xi_psi_N}, we can prove the following result.


\begin{proposition}
\label{prop:prop2}
For every $N \ge 1$ and every fixed realization of the amplitudes $\{Z_k\}_{k=1}^N$, the linear response prediction computed with the truncated noise $\xi^{(N)}$ coincides with the full linear response result~\eqref{eq:lin_response_result}, both for a single realization of the amplitudes and for the double expectation:
\begin{equation}
\hat{\mathbb{E}}^{\bar{f}_t}[q^2]^{(N)} = \mathbb{E}^{\bar{f}_t}[q^2]^{(N)} = \mathbb{E}^{\bar{f}_t}[q^2] = \frac{e^{-2\omega^2 t/\gamma}}{m\beta_0\omega^2}, \qquad \forall N \ge 1.
\label{eq:LRT_truncation}
\end{equation}
\end{proposition}
\noindent
In other words, a finite-order truncation of the K--L expansion does not alter the linear response prediction.

\begin{proof}
At $t = 0$ one has $\theta_0 = \cos\psi$, so that the initial value of the truncated noise reads
\begin{equation}
\xi_0^{(N)}(\psi) = \sqrt{\frac{2}{T}} \sum_{k=1}^{N} Z_k \cos\big((2k-1)\psi\big) = \sqrt{\frac{2}{T}} \sum_{k=1}^{N} Z_k\, T_{2k-1}(\theta_0),
\end{equation}
and, for a fixed realization of the amplitudes $\{Z_k\}_{k=1}^N$, the corresponding dissipation function is
\begin{equation}
\tilde{\Omega}_{(N)}^{\tilde{f}_0}(q_0, \psi) = \frac{\omega^2}{\gamma} - \frac{\omega^4}{\gamma}m\beta_0 q_0^2 + \omega^2 m\beta_0 \sigma\, q_0\, \xi_0^{(N)}(\psi).
\end{equation}
In the linear response formula~\eqref{eq:LRT_formula}, the observable evolves along the unperturbed flow, $q^2 \circ \tilde{S}_0^s = q_0^2 e^{-2\omega^2 s/\gamma}$. Since $q_0$ and $\psi$ are independent, with densities $f_0$ and $g_0(\psi) = (2\pi)^{-1}$, the expectation at fixed amplitudes factorizes as
\begin{align}
\hat{\mathbb{E}}^{\tilde{f}_0}\!\left[\tilde{\Omega}_{(N)}^{\tilde{f}_0} \left(q^2 \circ \tilde{S}_0^s\right)\right] &= \int_{0}^{2\pi} \mathrm{d}\psi \, g_0(\psi) \int_{-\infty}^{\infty} \mathrm{d}q_0 \, f_0(q_0) \, \tilde{\Omega}_{(N)}^{\tilde{f}_0}(q_0, \psi) \, q_0^2\, e^{-2\omega^2 s/\gamma} \nonumber \\
&= e^{-2\omega^2 s/\gamma} \left( \frac{\omega^2}{\gamma}\mathbb{E}^{f_0}[q_0^2] - \frac{\omega^4}{\gamma}m\beta_0\, \mathbb{E}^{f_0}[q_0^4] \right) \nonumber \\
&\quad + \omega^2 m\beta_0 \sigma\, e^{-2\omega^2 s/\gamma} \, \mathbb{E}^{f_0}[q_0^3] \, \sqrt{\frac{2}{T}} \sum_{k=1}^{N} Z_k\, \hat{\mathbb{E}}^{g_0}\!\left[ T_{2k-1}(\theta_0) \right].
\label{eq:lrt_integrand_chebyshev}
\end{align}
The noise-dependent term vanishes through two independent mechanisms:
\begin{itemize}
\item[(i)] the initial density $f_0$ is even, so that $\mathbb{E}^{f_0}[q_0^3] = 0$;
\item[(ii)] for every mode $k \ge 1$, the average over the uniform phase vanishes,
\begin{equation*}
\hat{\mathbb{E}}^{g_0}\!\left[ T_{2k-1}(\theta_0) \right] = \frac{1}{2\pi}\int_0^{2\pi} \cos\big((2k-1)\psi\big)\,\mathrm{d}\psi = 0.
\end{equation*}
\end{itemize}
Hence, for every realization of the amplitudes, using $\mathbb{E}^{f_0}[q_0^2] = (m\beta_0\omega^2)^{-1}$ and $\mathbb{E}^{f_0}[q_0^4] = 3(m\beta_0\omega^2)^{-2}$,
\begin{equation}
\hat{\mathbb{E}}^{\tilde{f}_0}\!\left[\tilde{\Omega}_{(N)}^{\tilde{f}_0} \left(q^2 \circ \tilde{S}_0^s\right)\right] = -\frac{2}{\gamma m \beta_0}\, e^{-2\omega^2 s/\gamma}.
\end{equation}
Inserting this result into Eq.~\eqref{eq:LRT_formula}, with $\mathbb{E}^{f_0}[q^2] = (m\beta_0\omega^2)^{-1}$, and integrating over $s \in [0,t]$ yields $\hat{\mathbb{E}}^{\bar{f}_t}[q^2]^{(N)} = e^{-2\omega^2 t/\gamma}/(m\beta_0\omega^2)$, i.e.\ the first equality in Eq.~\eqref{eq:LRT_truncation} together with the explicit value. Since this result does not depend on the amplitudes, averaging over them leaves it unchanged, which gives the second equality (the zero mean of the amplitudes, $\mathbb{E}_Z[Z_k] = 0$, 
provides a third, independent reason for the vanishing of the noise term). Finally, since the result is also independent of $N$, it coincides with its limit as $N$ grows without bounds.

\end{proof}

\section{Conclusions}
\label{sec:Concl}

We have applied Exact Response Theory to an overdamped Brownian oscillator driven by thermal white noise, using a K--L representation to embed the forcing in an autonomous augmented phase space. The purpose was to validate the applicability of ERT on a simple and exactly solvable model, with the  
intent of extending it well beyond such models. ERT could in fact be applied to non-Markovian processes, as well as in presence of phase transitions, not just under quite general forcings.

The construction makes the role of the auxiliary variables explicit: for fixed spectral amplitudes, the oscillator and the harmonic phase evolve deterministically, while the stochastic expectation is recovered by averaging over the amplitudes and the initial phase. In this setting, the dissipation function can be used in the ERT response formula even though the original model is given as a stochastic differential equation. Uniform sampling of the initial auxiliary phase is essential: it removes phase-dependent contributions from the noise correlation and produces the autocorrelation needed in the white-noise limit. This requirement is part of the ensemble construction, rather than a dispensable choice of initial condition.

For the second moment $q^2$, the response obtained with a finite number $N$ of K--L modes admits a closed form for every $N$. As $N$ grows, it approaches the generator-based result, and in the limit $N\to\infty$ followed by $T\to\infty$ it recovers exactly the known relaxation from the initial variance to the thermal variance of the bath, Eq.~\eqref{eq:generator_target}, and Proposition~\ref{prop:prop1}.

The comparison with linear response clarifies why the exact calculation is needed, and reminds us that the performance of a response theory obviously depends on the observable of interest. At first order, the noise contributions vanish for every observable depending on $q$ only, so that linear response retains only the deterministic relaxation of the initial density. The variance injected by the bath is proportional to the square of the noise amplitude, and therefore lies beyond any first-order approximation. For the observable $q^2$, linear response predicts an exponential decay to zero, whereas ERT reproduces the full variance relaxation, including the approach to the steady-state value dictated by the thermal bath.

The calculations reported above already contain more than has been used here. The exact response
formula \eqref{response_formula} and its integrand \eqref{eq:cross_term_integral} are not tied to any particular statistical interpretation of the
additional variables: what changes, as more of them are averaged over, is only the correlation of the
noise entering \eqref{eq:cross_term_integral}, which passes from the empirical correlation of a single realization of the
forcing, at fixed amplitudes and fixed phase, to the expression obtained by averaging over the
phase, 
and finally to \eqref{eq:CT_N}, the
average over the amplitudes merely replacing the squares $Z_k^2$ by their mean value. The closed
form \eqref{eq:q2_N} accordingly holds at fixed amplitudes with $Z_k^2$ in place of unity, so that for this
observable a deterministic forcing may reproduce the stochastic response exactly. The two sets of
variables behave differently in this respect. The amplitudes are conserved by the
augmented dynamics, and their law may be prescribed at will, concentrated as in Proposition 
\ref{prop:prop2} or
Gaussian as in Section \ref{sec:limit}, without altering the dissipation function; the phase is not conserved,
and its distribution enters the last term of \eqref{dissipation}. The uniform choice adopted here is the invariant
one, for which that term vanishes and the pure noise contribution to the response vanishes
identically, Eq.~\eqref{eq:term3_eval} of Appendix \ref{app:appC}, while any other distribution with full support is
admissible, and would describe a perturbation that is known with finite accuracy, when it is switched on.

This hierarchy of descriptions is invisible to linear response, the three mechanisms that make the
first order noise contribution vanish in Proposition \ref{prop:prop2} belonging one to each of its levels, whereas
the exact response distinguishes them. Its systematic development, together with that of the role
of $g_0$, is left to future work.

To compare with the existing stochastic and deterministic approaches, considered separately, we note that
generator is defined only
for Markov processes, whereas the Karhunen--Lo\`eve representation uses nothing but the covariance
of the forcing, which exists irrespective of the Markov property.
The response identity holds at fixed
amplitudes and fixed phase, so that the ensemble over the realizations of the noise is built
explicitly, rather than presupposed as in the forward Kolmogorov equation \eqref{forwop}, whose solution is an
average over the noise from the outset. Correspondingly the dissipation function retains the form
\eqref{11} of a phase space quantity evolving along a deterministic flow, while the quantity playing the
same role in the generator based formulation, Eq.~\eqref{Om}, involves the diffusion matrix and the second
derivatives of the density. 
Also, the finite truncations, moreover, provide a family of closed form
approximants of the noise that
are interesting of their own merits. Finally, no invariance of $f_0$ under the unperturbed
dynamics is required here, although it may be needed for specific applications.

In the theory of deterministic
dynamical systems one averages over the initial conditions of the system at a fixed forcing, in the
theory of stochastic processes one averages over the realizations of the forcing at a given initial
condition, and the augmented formulation accommodates either average, as well as their composition.
The last keeps the indetermination of the state of the system, carried by $f_0$, and that of
the perturbation, carried by $g_0$ and by the law of the amplitudes, as separate and independently
assignable objects, so that the fluctuations of the response may be apportioned between the two.
Neither the state of the system nor the perturbation can ever ever be 
specified with infinite precision, and both are accordingly described here by distributions of
finite width. The calculation of Section \ref{sec:LRT}
is an instance of what this entails.

The present analysis is restricted to a linear oscillator and its second moment. For finite $T$ and a finite number of modes, the forcing is only an approximation to white noise; convergence to the response under white noise requires controlling these limits, whose order matters: the number of modes must diverge before the time horizon does. We also remark that while our calculation does not establish that every stochastic or non-Markovian process admits the same augmented-space treatment, it provides, however, a meaningful starting point for such extensions. Natural next cases that we plan to address in future works include 
colored noise with a given covariance, a systematic treatment of the
levels of description and of the role of $g_0$, additional observables, and nonlinear or
multidimensional systems. In all these cases, the structure of the augmented phase space and the convergence of the response averages as the truncation order increases will need to be assessed case by case..




\appendix

\section{The Karhunen-Loève theorem}
\label{app:appA}

In this Appendix section, we briefly recall the K--L theorem, which allows a stochastic process to be represented as an infinite linear combination of orthogonal functions, analogous to a Fourier series, utilizing stochastic rather than deterministic coefficients. If $\{X_t\}_{t \in [a,b]}$ is a centered stochastic process satisfying suitable continuity conditions, it can be decomposed into a sum of pairwise orthogonal functions multiplied by random coefficients that are pairwise uncorrelated (and hence, orthogonal in a suitable functional space). More precisely, the following holds \cite{Loeve}:
\begin{Theorem}[Karhunen--Loève \cite{pavliotis2014stochastic}]
Let $X_t$, $t \in [a,b]$ be a square-integrable stochastic process with zero mean, defined on a probability space ($\Theta, F, P)$, with a continuous covariance function $K_X(s,t)$. By letting $e_k$ be an orthonormal basis on ${L}^2([a,b])$ formed by the eigenfunctions of $T_{K_X}$ with respective eigenvalues $\lambda_k$, $X_t$ admits the following representation:
\begin{equation}
X_t= \sum_{k=1}^\infty Z_k e_k(t),
\end{equation}
where the convergence is in ${L}^2$, uniform in $t$, and:
\begin{equation}
    Z_k=\int_a^b X_t e_k(t)dt.
\end{equation}
Furthermore, the random variables $Z_k$ have a zero mean, are uncorrelated, and possess variance $\lambda_k$:
\begin{equation}
    \langle Z_k\rangle=0, \hspace{0.5cm} \forall k \in \mathbb{N} \quad \text{and} \quad \langle Z_i Z_j \rangle=\delta_{ij}\lambda_j, \hspace{0.5cm} \forall i,j \in \mathbb{N}.
\end{equation}
\end{Theorem}

As an example, consider the particular case of a standard Wiener process, $w_t$, for $t \in [0,T]$, with covariance function $K_w(t,s)=\min(s,t)$. To find the corresponding eigenvalues and eigenfunctions, we solve the following integral equation:
\begin{equation}
    \int_a^b K_w(s,t)e(s)ds = \int_0^T \min(s,t)e(s)ds = \int_0^t se(s)ds + t\int_t^T e(s)ds = \lambda e(t) \, ,\hspace{0.5cm} 0\leq t \leq T
\end{equation}
Differentiating once with respect to $t$ yields:
\begin{equation}
     \int_t^Te(s)ds=\lambda e'(t)
\label{firstD}
\end{equation}
A second differentiation produces $-e(t)=\lambda e''(t)$, whose general solution takes the harmonic form:
\begin{equation}
     e(t)=A\sin{\left( \frac{t}{\sqrt{\lambda}}\right)}+B\cos{\left( \frac{t}{\sqrt{\lambda}}\right)},
\end{equation}
where $A$ and $B$ are determined by the boundary conditions. Setting $t=0$ in the integral equation gives $e(0)=0$, implying $B=0$. Setting $t=T$ in Equation \eqref{firstD} yields $e'(T)=0$, which requires:
\begin{equation}
\lambda_k=\left( \frac{T}{(k-\frac{1}{2})\pi} \right)^2, \hspace{1cm} k\geq1.
\end{equation}
The corresponding eigenfunctions are thus:
\begin{equation}
    e_k(t)=A\sin{\left[(k-\frac{1}{2})\pi \frac{t}{T} \right]}, \hspace{1cm} k\geq1 
\end{equation}
The amplitude $A$ is determined by the normalization condition $\int_0^T e^2_k(t)dt = 1$, which leads to $A=\sqrt{\frac{2}{T}}$. Finally, we obtain the K--L representation of the Wiener process:
\begin{equation}
W(t)=\frac{\sqrt{2T}}{\pi} \sum_{k=1}^\infty Z_k \frac{\sin{\left[(k-\frac{1}{2})\pi \frac{t}{T}\right]}}{(k-\frac{1}{2})},
\label{KLexp}
\end{equation}
where $\{Z_k\}_{k=1}^\infty$ is a sequence of independent Gaussian random variables, each having a zero mean and unit variance. This representation is valid for $t\in[0,T]$ for any $T>0$, and convergence is uniform in $t$ with respect to the ${L}^2$ norm.

\section{Expansion in Chebyshev polynomials}
\label{app:appB}

In this Appendix section we provide the derivation of the explicit Chebyshev series used to rewrite the multi-frequency thermal noise in terms of the single augmented phase space coordinate $\theta(t)$. We subsequently demonstrate how evaluating  the time evolution of the series algebraically collapses the expansion back into the standard explicit time-dependent representation.

Using De Moivre's identity we can write:
\begin{equation}
\cos(n\alpha) + i\sin(n\alpha) = (\cos\alpha + i\sin\alpha)^n
\end{equation}
We expand the right-hand side using the binomial theorem:
\begin{equation}
(\cos\alpha + i\sin\alpha)^n = \sum_{j=0}^{n} \binom{n}{j} (\cos\alpha)^{n-j} (i\sin\alpha)^j
\end{equation}
Since we are evaluating a pure cosine function, we isolate the real part of this expansion. The real terms correspond  to the even indices $j = 2p$, where the imaginary unit evaluates to $i^{2p} = (-1)^p$:
\begin{equation}
\cos(n\alpha) = \sum_{p=0}^{ [ n/2 ]} \binom{n}{2p} (-1)^p (\cos\alpha)^{n-2p} (\sin^2\alpha)^p
\label{cos_series}
\end{equation}
Using $\sin^2{\alpha}+\cos^2{\alpha}=1$ and letting $\theta \equiv \cos\alpha$, we obtain the general form of the Chebyshev polynomial of the first kind \cite{mason2002chebyshev}, defined as $T_n(\cos{\alpha})=\cos{(n\alpha)}$, so that:
\begin{equation}
T_n(\theta) = \sum_{p=0}^{ [n/2] } \binom{n}{2p} (-1)^p \theta^{n-2p} (1 - \theta^2)^p
\label{Tn}
\end{equation}
By further expanding the binomial $(1 - \theta^2)^p$ and collecting terms of equal powers of $\theta$, the polynomial takes its standard explicit series representation in descending powers:
\begin{equation}
T_n(\theta) = \frac{n}{2} \sum_{j=0}^{ [n/2] } (-1)^j \frac{(n-j-1)!}{j!(n-2j)!} (2\theta)^{n-2j}
\label{Tn2}
\end{equation}
It is convenient to rewrite formula~\eqref{Tn2} writing $n = 2k-1$, which leads to:
\begin{equation}
T_{2k-1}(\theta) = \left(2k-1\right) \sum_{j=0}^{k-1} (-1)^j \frac{(2k-j-2)!}{j!(2k-2j-1)!} 2^{2k-2j-2} \theta^{2k-2j-1}
\end{equation}
Substituting this expression into the K--L expansion \eqref{eq:xi_psi} yields the following explicit expression in $\theta$ for the white noise:
\begin{equation}
\xi(\theta) = \sqrt{\frac{2}{T}} \sum_{k=1}^{N} Z_k \left[ \left(2k-1\right) \sum_{j=0}^{k-1} (-1)^j \frac{(2k-j-2)!}{j!(2k-2j-1)!} 2^{2k-2j-2} \theta^{2k-2j-1} \right]
\label{eq:full_double_series}
\end{equation}

The time evolution of the augmented variable is $\theta(t) = \cos(bt + \psi)$. By definition of the De Moivre derivation above, substituting a cosine into the explicit Chebyshev power series forces the entire  summation to collapse into a single cosine, to see that it is sufficient to perform the last computations from \eqref{cos_series} to \eqref{eq:full_double_series}  backwards replacing $\theta$ with $\theta(t)$:
\begin{equation}
T_{2k-1}(\cos(bt + \psi)) = \cos((2k-1)(bt + \psi)).
\end{equation}
Therefore, evaluating the time evolution of the noise yields:
\begin{align}
\xi(t) &= \sqrt{\frac{2}{T}} \sum_{k=1}^{N} Z_k \cos((2k-1)(bt + \psi)) \nonumber \\
&= \sqrt{\frac{2}{T}} \sum_{k=1}^{N} Z_k \cos((2k-1)bt + (2k-1)\psi).
\end{align}
Recalling our parameterization for the fundamental frequency $b = \frac{\pi}{2T}$ and the base phase $\psi$, we explicitly recover the defining relations for the K--L spectrum, namely $b_k = (2k -1)b$ and $\psi_k=(2k-1) \psi$. This restores the standard explicit time-dependent representation of the white noise:
\begin{equation}
\xi(t) = \sqrt{\frac{2}{T}} \sum_{k=1}^{N} Z_k \cos(b_k t + \psi_k) .
\label{xiKL}
\end{equation}

\section{Response formula}
\label{app:appC}

To derive the exact response integrand $\mathbb{E}^{\tilde{f}_0}\!\left[\tilde{\Omega}^{\tilde{f}_0}q^2(s)\right]$ in Eq.~\eqref{eq:cross_term_integral}, we evaluate the joint expectation of the dissipation function $\tilde{\Omega}^{\tilde{f}_0}$ and the squared trajectory $q^2(s)$. Throughout this appendix, expectations follow the convention of Section~\ref{sec:stochper}; in particular, $\mathbb{E}^{g_0}[\cdot]$ denotes the double expectation over the initial phase $\psi$ and the amplitudes $\{Z_k\}_{k\ge1}$, which are independent of $q_0$.

The solution of the perturbed overdamped Langevin equation $\dot{q} = -\frac{\omega^2}{\gamma}q + \sigma \xi(t)$ with initial position $q(0) = q_0$ reads
\begin{equation}
q(s) = q_0 e^{-\frac{\omega^2}{\gamma}s} + \sigma I(s), \qquad I(s) := \int_{0}^{s} e^{-\frac{\omega^2}{\gamma}(s-\tau)} \xi(\tau) \, \mathrm{d}\tau.
\label{eq:app_I_def}
\end{equation}
While this descends directly from the theory of stochastic processes \cite{Billingsley1995}, it is instructive to derive it explicitly in our setting.
Let us address first the convergence of the noise integral in Eq.~\eqref{eq:app_I_def}. For a finite truncation of the series~\eqref{eq:xi_psi}, $I(s)$ is an ordinary integral of a smooth function. For the full series, we define $I(s) := \lim_{N\to\infty} I_N(s)$, where $I_N(s)$ is obtained by integrating the first $N$ modes term by term. Writing $a := \omega^2/\gamma$, one has
\begin{equation*}
I_N(s) = \sqrt{\frac{2}{T}}\sum_{k=1}^{N} Z_k\, c_k(s),
\end{equation*}
with
\begin{equation*}
c_k(s) = \frac{a\cos(b_k s+\psi_k) + b_k\sin(b_k s+\psi_k) - e^{-as}\big(a\cos\psi_k + b_k\sin\psi_k\big)}{a^2+b_k^2}.
\end{equation*}
By the Cauchy--Schwarz inequality, $|a\cos x + b_k\sin x| \le \sqrt{a^2+b_k^2}$ for every $x$, which gives $|c_k(s)| \le 2(a^2+b_k^2)^{-1/2}$ for all $s \ge 0$ and all $\psi$. Therefore, using $\mathbb{E}_Z[Z_j Z_k] = \delta_{jk}$, for $N > M$, one finds
\begin{equation*}
\mathbb{E}_Z\big[(I_N(s) - I_M(s))^2\big] = \frac{2}{T}\sum_{k=M+1}^{N} c_k(s)^2 \le \frac{8}{T}\sum_{k>M}\frac{1}{a^2+b_k^2} \xrightarrow[M\to\infty]{} 0.
\end{equation*}
Since the right-hand side does not depend on $N$ and is the tail of a convergent series, $\{I_N(s)\}_{N\ge1}$ is a Cauchy sequence in mean square. By the completeness of the space of square-integrable random variables, it converges in mean square to a limit, which defines $I(s)$ \cite[Section 19]{Billingsley1995}. As a mean-square limit of centered Gaussian variables, $I(s)$ is, for fixed $\psi$, a centered Gaussian random variable, and, using $\sum_{k\ge1}(a^2+b_k^2)^{-1} = \frac{T}{2a}\tanh(aT)$ (see the footnote in Sec.~\ref{sec:trunc}), its variance satisfies $\mathbb{E}_Z[I(s)^2] \le 4\tanh(aT)/a$, uniformly in $s$ and $\psi$. Consequently, $q(s)$ and $q^2(s)$ are well-defined random variables.
Squaring $q(s)$ yields three distinct components:
\begin{equation}
q^2(s) = q_0^2 e^{-2\frac{\omega^2}{\gamma}s} + 2 \sigma q_0 e^{-\frac{\omega^2}{\gamma}s} I(s) + \sigma^2 I^2(s).
\label{eq:q2_decomposition}
\end{equation}
Recall that the initial dissipation function~\eqref{omega_definition} reads
\begin{equation}
\tilde{\Omega}^{\tilde{f}_0}(q_0,\theta_0,\phi_0) = \frac{\omega^2}{\gamma} - \frac{\omega^4}{\gamma}m\beta_0 q_0^2 + \omega^2 m\beta_0 \sigma\, q_0\, \xi_0.
\label{eq:dissipation_ref}
\end{equation}
Unlike $I(s)$, the initial noise $\xi_0$ is not defined pointwise for the full series; expectations involving it are understood term by term, i.e.\ as limits of the corresponding truncated expressions. Since $\tilde{f}_0 = f_0\, g_0$ factorizes and $f_0$ is a centered Gaussian distribution at parameter $\beta_0$, all odd moments of $q_0$ vanish, $\mathbb{E}^{f_0}[q_0] = \mathbb{E}^{f_0}[q_0^3] = 0$, while $\mathbb{E}^{f_0}[q_0^2] = (m\beta_0\omega^2)^{-1}$ and $\mathbb{E}^{f_0}[q_0^4] = 3(m\beta_0\omega^2)^{-2}$.

We now multiply Eq.~\eqref{eq:dissipation_ref} by each term in Eq.~\eqref{eq:q2_decomposition} and compute their expectations under $\mathbb{E}^{\tilde{f}_0}[\cdot]$.

\paragraph{1. Deterministic term.}
Multiplying $\tilde{\Omega}^{\tilde{f}_0}$ by the first term $q_0^2 e^{-2\frac{\omega^2}{\gamma}s}$ gives
\begin{align}
\mathbb{E}^{\tilde{f}_0}\!\left[ \tilde{\Omega}^{\tilde{f}_0}\, q_0^2 e^{-2\frac{\omega^2}{\gamma}s} \right] &= e^{-2\frac{\omega^2}{\gamma}s} \,\mathbb{E}^{f_0}\!\left[ \frac{\omega^2}{\gamma} q_0^2 - \frac{\omega^4}{\gamma}m\beta_0 q_0^4 \right] + \omega^2 m\beta_0 \sigma\, e^{-2\frac{\omega^2}{\gamma}s} \,\mathbb{E}^{f_0}[q_0^3]\, \mathbb{E}^{g_0}[\xi_0] \nonumber \\
&= e^{-2\frac{\omega^2}{\gamma}s} \left( \frac{\omega^2}{\gamma} \frac{1}{m\beta_0\omega^2} - \frac{\omega^4}{\gamma}m\beta_0 \frac{3}{(m\beta_0\omega^2)^2} \right) = -\frac{2}{\gamma m \beta_0}\, e^{-2\frac{\omega^2}{\gamma}s},
\label{eq:term1_eval}
\end{align}
where the last term in the first line vanishes since $\mathbb{E}^{f_0}[q_0^3] = 0$.

\paragraph{2. Cross term.}
Multiplying $\tilde{\Omega}^{\tilde{f}_0}$ by the second term $2 \sigma q_0 e^{-\frac{\omega^2}{\gamma}s} I(s)$ yields
\begin{align}
\mathbb{E}^{\tilde{f}_0}\!\left[ \tilde{\Omega}^{\tilde{f}_0}\, 2 \sigma q_0 e^{-\frac{\omega^2}{\gamma}s} I(s) \right] &= 2\sigma e^{-\frac{\omega^2}{\gamma}s}\, \mathbb{E}^{f_0}\!\left[ \frac{\omega^2}{\gamma}q_0 - \frac{\omega^4}{\gamma}m\beta_0 q_0^3 \right] \mathbb{E}^{g_0}[I(s)] \nonumber \\
&\quad + 2 \sigma^2 e^{-\frac{\omega^2}{\gamma}s}\, \omega^2 m\beta_0\, \mathbb{E}^{f_0}\!\left[ q_0^2 \right] \mathbb{E}^{g_0}[\xi_0 I(s)].
\end{align}
The first expectation over $f_0$ vanishes, since it involves only odd powers of $q_0$. Using $\mathbb{E}^{f_0}[q_0^2] = (m\beta_0\omega^2)^{-1}$ in the second one, and exchanging the expectation with the time integral defining $I(s)$, which is legitimate for every truncation and carries over to the full series term by term, we obtain
\begin{equation}
\mathbb{E}^{\tilde{f}_0}\!\left[ \tilde{\Omega}^{\tilde{f}_0}\, 2 \sigma q_0 e^{-\frac{\omega^2}{\gamma}s} I(s) \right] = 2 \sigma^2 e^{-\frac{\omega^2}{\gamma}s}\, \mathbb{E}^{g_0}[\xi_0 I(s)] = 2 \sigma^2 e^{-\frac{\omega^2}{\gamma}s} \int_{0}^{s} e^{-\frac{\omega^2}{\gamma}(s-\tau)}\, C_T(\tau) \, \mathrm{d}\tau,
\label{eq:term2_eval}
\end{equation}
with $C_T$ the noise autocorrelation~\eqref{eq:CT}. The evaluation of the remaining integral, in which $C_T$ is concentrated at the endpoint $\tau = 0$ in the large-$T$ limit, is discussed in Section~\ref{sec:limit} and Appendix~\ref{app:appD}.

\paragraph{3. Pure noise term.}
Multiplying $\tilde{\Omega}^{\tilde{f}_0}$ by the third term $\sigma^2 I^2(s)$ gives
\begin{equation}
\mathbb{E}^{\tilde{f}_0}\!\left[ \tilde{\Omega}^{\tilde{f}_0}\, \sigma^2 I^2(s) \right] = \sigma^2\, \mathbb{E}^{f_0}\!\left[ \frac{\omega^2}{\gamma} - \frac{\omega^4}{\gamma}m\beta_0 q_0^2 \right] \mathbb{E}^{g_0}[I^2(s)] + \omega^2 m\beta_0 \sigma^3\, \mathbb{E}^{f_0}[q_0]\, \mathbb{E}^{g_0}[\xi_0 I^2(s)],
\end{equation}
where $\mathbb{E}^{g_0}[I^2(s)]$ is finite by the bound above. The second term vanishes, since $\mathbb{E}^{f_0}[q_0] = 0$ (for every truncation, and hence for the full series), while the expectation over $f_0$ in the first term evaluates to
\begin{equation}
\mathbb{E}^{f_0}\!\left[ \frac{\omega^2}{\gamma} - \frac{\omega^4}{\gamma}m\beta_0 q_0^2 \right] = \frac{\omega^2}{\gamma} - \frac{\omega^4}{\gamma}m\beta_0\, \frac{1}{m\beta_0\omega^2} = 0.
\end{equation}
Thus, the contribution of the pure noise term vanishes identically:
\begin{equation}
\mathbb{E}^{\tilde{f}_0}\!\left[ \tilde{\Omega}^{\tilde{f}_0}\, \sigma^2 I^2(s) \right] = 0.
\label{eq:term3_eval}
\end{equation}
Combining Eqs.~\eqref{eq:term1_eval}, \eqref{eq:term2_eval} and~\eqref{eq:term3_eval} yields Eq.~\eqref{eq:cross_term_integral}:
\begin{equation*}
\mathbb{E}^{\tilde{f}_0}\!\left[\tilde{\Omega}^{\tilde{f}_0}q^2(s)\right] = -\frac{2}{\gamma m \beta_0}\, e^{-2\frac{\omega^2}{\gamma}s} + 2 \sigma^2 e^{-\frac{\omega^2}{\gamma}s} \int_{0}^{s} e^{-\frac{\omega^2}{\gamma}(s-\tau)}\, C_T(\tau) \, \mathrm{d}\tau.
\end{equation*}

\section{Noise autocorrelation and heuristic large-$T$ limit}
\label{app:appD}

We first compute the truncated autocorrelation $C_T^{(N)}(\tau) = \mathbb{E}^{g_0}\big[\xi^{(N)}(\tau)\,\xi^{(N)}_0\big]$, where, as in Section~\ref{sec:stochper}, $\mathbb{E}^{g_0}[\cdot]$ denotes the double expectation over both the initial phase $\psi$ and the amplitudes $\{Z_k\}_{k\ge1}$. Using $\mathbb{E}_Z[Z_k Z_j] = \delta_{kj}$, the double sum collapses into a single sum:
\begin{equation}
\begin{aligned}
\mathbb{E}_Z\big[\xi^{(N)}(\tau)\,\xi^{(N)}_0\big] &= \frac{2}{T} \sum_{k=1}^{N} \sum_{j=1}^{N}\mathbb{E}_Z[Z_k Z_j]\, \cos(b_k \tau + \psi_k) \cos\psi_j \\
&= \frac{2}{T} \sum_{k=1}^{N} \cos(b_k \tau + \psi_k) \cos\psi_k.
\end{aligned}
\end{equation}
By the identity $\cos A \cos B = \frac{1}{2}\left[\cos(A - B) + \cos(A + B)\right]$, each term splits into a phase-independent and a phase-dependent part,
\begin{equation}
\cos(b_k \tau + \psi_k) \cos\psi_k = \frac{1}{2} \cos(b_k \tau) + \frac{1}{2} \cos\big(b_k \tau + m_k\psi\big), \qquad m_k = 2(2k - 1),
\label{103}
\end{equation}
where we used $\psi_k = (2k-1)\psi$. Since $m_k$ is a non-zero integer, the phase-dependent part vanishes upon averaging over $\psi \sim \mathcal{U}[0, 2\pi)$:
\begin{equation}
\frac{1}{2\pi} \int_{0}^{2\pi} \cos\big(b_k \tau + m_k \psi\big) \, \mathrm{d}\psi = \frac{1}{2\pi m_k} \Big[ \sin(b_k \tau + 2\pi m_k) - \sin(b_k \tau) \Big] = 0
\label{104}
\end{equation}
Hence
\begin{equation}
C_T^{(N)}(\tau) = \frac{1}{T} \sum_{k=1}^{N} \cos(b_k \tau),
\end{equation}
which is Eq.~\eqref{eq:CT_N}. The full autocorrelation $C_T$ is obtained in the limit $N\to\infty$, understood term by term, which gives Eq.~\eqref{eq:CT_riemann}.

We now detail the heuristic argument of Section~\ref{sec:limit}. As $T \to \infty$, the spacing $\Delta b = \pi/T$ vanishes and the lowest frequency $b_1 = \pi/(2T)$ tends to zero, so that the Riemann sum~\eqref{eq:CT_riemann} formally converges to an integral over $\nu \in [0,\infty)$. By the even symmetry of the cosine, and since the sine component integrates to zero over a symmetric interval,
\begin{equation}
\lim_{T \to \infty} C_T(\tau) = \frac{1}{\pi} \int_{0}^{\infty} \cos(\nu \tau) \, \mathrm{d}\nu = \frac{1}{2\pi} \int_{-\infty}^{\infty} e^{i \nu \tau} \, \mathrm{d}\nu = \delta(\tau),
\label{formal}
\end{equation}
where the equalities hold in the distributional sense. 
Inserting this result into Eq.~\eqref{eq:cross_term_integral}, i.e.\ formally taking the limit $T\to\infty$ in the integrand, and noting that a Dirac delta located at the lower boundary of the integration domain contributes with weight $1/2$,
\begin{equation*}
\int_{0}^{s} e^{-\frac{\omega^2}{\gamma}(s-\tau)} \delta(\tau) \, \mathrm{d}\tau = \frac{1}{2} e^{-\frac{\omega^2}{\gamma}s},
\end{equation*}
we obtain
\begin{equation}
\mathbb{E}^{\tilde{f}_0}\!\left[\tilde{\Omega}^{\tilde{f}_0}q^2(s)\right] = -\frac{2}{\gamma m \beta_0} e^{-2\frac{\omega^2}{\gamma}s} + \sigma^2 e^{-2\frac{\omega^2}{\gamma}s} = \left( \sigma^2 - \frac{2}{\gamma m \beta_0} \right) e^{-2\frac{\omega^2}{\gamma}s}.
\label{eq:dissipation_simplified}
\end{equation}
Inserting Eq.~\eqref{eq:dissipation_simplified} into the response formula~\eqref{response_formula} and integrating over $s \in [0, t]$ gives
\begin{equation}
\mathbb{E}^{\tilde{f}_t}[q^2] = \frac{1}{m\beta_0\omega^2} + \left( \sigma^2 - \frac{2}{\gamma m \beta_0} \right) \frac{\gamma}{2\omega^2} \left( 1 - e^{-2\frac{\omega^2}{\gamma}t} \right).
\end{equation}
Finally, using $\sigma^2 = 2(\gamma m \beta)^{-1}$,
\begin{align}
\mathbb{E}^{\tilde{f}_t}[q^2] &= \frac{1}{m\beta_0\omega^2} + \left( \frac{1}{m\beta\omega^2} - \frac{1}{m\beta_0\omega^2} \right) \left( 1 - e^{-2\frac{\omega^2}{\gamma}t} \right) \nonumber \\
&= \frac{e^{-2\frac{\omega^2}{\gamma}t}}{m\beta_0\omega^2} + \frac{1 - e^{-2\frac{\omega^2}{\gamma}t}}{m\beta\omega^2},
\end{align}
which coincides with Eq.~\eqref{eq:generator_target}. The rigorous counterpart of this computation is provided in the Proposition \ref{prop:prop1} of Section~\ref{sec:limit}.

\section{Derivation of the linear response formula}
\label{app:appE}

In this appendix, we derive the linear response formula~\eqref{eq:LRT_formula} as the first-order expansion, in the perturbation parameter $\sigma$, of the exact response, for observables depending on $q$ only. Throughout, we set $a = \omega^2/\gamma$ and work at a fixed realization of the amplitudes $\{Z_k\}_{k\ge1}$; the corresponding double expectations follow by a further average over the amplitudes.

We observe that the dissipation function~\eqref{omega_definition} is affine in $\sigma$:
\begin{equation}
\tilde{\Omega}^{\tilde{f}_0}(q_0,\theta_0,\phi_0) = \tilde{\Omega}_0(q_0) + \sigma\,\tilde{\Omega}_1(q_0,\theta_0),
\label{eq:app_Omega_split}
\end{equation}
with
\begin{equation}
\tilde{\Omega}_0(q_0) = \frac{\omega^2}{\gamma} - \frac{\omega^4}{\gamma}m\beta_0 q_0^2,
\qquad
\tilde{\Omega}_1(q_0,\theta_0) = \omega^2 m\beta_0\, q_0\, \xi_0(\theta_0).
\label{eq:app_Omega01}
\end{equation}
The unperturbed part $\tilde{\Omega}_0$ is the dissipation function of the deterministic flow $\dot{q} = -a q$ with respect to $f_0$. We point out that it does not vanish, since $f_0$ is not invariant for this flow, but it has zero mean, $\mathbb{E}^{f_0}[\tilde{\Omega}_0] = a - a\, m\beta_0\omega^2\,\mathbb{E}^{f_0}[q_0^2] = 0$. All the noise dependence is carried by $\tilde{\Omega}_1$, which is odd in $q_0$.
Keeping the notation of App.~\ref{app:appC}, we write
\begin{equation*}
I(s) := \int_0^s e^{-a(s-\tau)}\,\xi(\tau)\,\mathrm{d}\tau, \qquad s \in \mathbb{R}.
\end{equation*}
We recall, from Eq.~\eqref{solq}, that for every $s \in \mathbb{R}$ the evolution of the observable $q$ along the perturbed flow obeys
\begin{equation}
q(s) = q_0\,e^{-as} + \sigma\,I(s),
\label{eq:app_traj}
\end{equation}
where $q_0 e^{-as}$ is the contribution of the unperturbed flow $\tilde{S}^s_0$ and $I(s)$ is evaluated along the auxiliary trajectory through $(\theta_0,\phi_0)$.
The auxiliary variables $(\theta,\phi)$ evolve independently of $\sigma$.

For a fixed realization of the amplitudes, the density evolved by the perturbed flow is given by:
\begin{equation}
\tilde{f}_t(\tilde{\Gamma}) = \exp\left\{\int_{-t}^{0}\tilde{\Omega}^{\tilde{f}_0}\big(\tilde{S}^s\tilde{\Gamma}\big)\,\mathrm{d}s\right\}\tilde{f}_0(\tilde{\Gamma}).
\label{eq:app_density}
\end{equation}
Note that the parameter $\sigma$ enters the exponent of Eq.~\eqref{eq:app_density} in two ways: explicitly, through $\sigma\,\tilde{\Omega}_1$, and implicitly, through the perturbed trajectory encoded in the flow $\tilde{S}^s$. By Eq.~\eqref{eq:app_traj}, with $q_0$ replaced by $q$, the $q$-component of $\tilde{S}^s\tilde{\Gamma}$ is $q\,e^{-as} + \sigma I(s)$, with $I$ evaluated along the auxiliary trajectory through $(\theta,\phi)$. A Taylor expansion of $\tilde{\Omega}_0$ around the unperturbed trajectory leads to:
\begin{equation}
\tilde{\Omega}^{\tilde{f}_0}\big(\tilde{S}^s\tilde{\Gamma}\big) = \tilde{\Omega}_0\big(\tilde{S}^s_0\tilde{\Gamma}\big) + \sigma\Big[\tilde{\Omega}_1\big(\tilde{S}^s_0\tilde{\Gamma}\big) + \tilde{\Omega}_0'\big(q\,e^{-as}\big)\,I(s)\Big] + O(\sigma^2),
\label{eq:app_Omega_expansion}
\end{equation}
where $\tilde{\Omega}_0'(x) = -2a\,m\beta_0\omega^2\,x$. The square bracket contains an explicit contribution, the perturbation $\tilde{\Omega}_1$ evaluated along the unperturbed trajectory, and an implicit one, the change of $\tilde{\Omega}_0$ caused by the displacement $\sigma I(s)$ of the trajectory. Inserting Eq.~\eqref{eq:app_Omega_expansion} into Eq.~\eqref{eq:app_density} and using $e^{A_0 + \sigma A_1} = e^{A_0}(1 + \sigma A_1) + O(\sigma^2)$, we obtain
\begin{equation}
\tilde{f}_t(\tilde{\Gamma}) = \tilde{f}^{(0)}_t(\tilde{\Gamma})\left[1 + \sigma\int_{-t}^{0}\Big(\tilde{\Omega}_1\big(\tilde{S}^s_0\tilde{\Gamma}\big) + \tilde{\Omega}_0'\big(q\,e^{-as}\big)\,I(s)\Big)\mathrm{d}s\right] + O(\sigma^2),
\label{eq:app_density_first_order}
\end{equation}
where
\begin{equation}
\tilde{f}^{(0)}_t(\tilde{\Gamma}) := \exp\left\{\int_{-t}^{0}\tilde{\Omega}_0\big(\tilde{S}^s_0\tilde{\Gamma}\big)\,\mathrm{d}s\right\}\tilde{f}_0(\tilde{\Gamma})
\label{eq:app_density_zero}
\end{equation}
is the density evolved under the unperturbed dynamics (recall, from Sec.~\ref{sec:ERform}, that in our framework $f_0$ is not invariant under this dynamics).

The zeroth-order density can be computed explicitly. Since $\tilde{\Omega}_0(q\,e^{-as}) = a - a\,m\beta_0\omega^2 q^2 e^{-2as}$,
\begin{equation*}
\int_{-t}^{0}\tilde{\Omega}_0\big(\tilde{S}^s_0\tilde{\Gamma}\big)\,\mathrm{d}s = at - \frac{m\beta_0\omega^2 q^2}{2}\big(e^{2at} - 1\big),
\end{equation*}
and, since $g_0$ is invariant under the oscillator dynamics,
\begin{equation}
\tilde{f}^{(0)}_t(q,\theta,\phi) = \sqrt{\frac{m\beta_0\omega^2 e^{2at}}{2\pi}}\,\exp\left(-\frac{m\beta_0\omega^2 e^{2at}}{2}\,q^2\right) g_0(\theta,\phi).
\label{eq:app_density_zero_liouville}
\end{equation}
This is a centered Gaussian of variance $e^{-2at}/(m\beta_0\omega^2)$, i.e.\ the initial distribution contracted by the deterministic relaxation.
The first-order term also takes a simple form. Using Eq.~\eqref{eq:app_Omega01} and $\tilde{\Omega}_0'(q\,e^{-as}) = -2a\,m\beta_0\omega^2 q\,e^{-as}$, the integrand in Eq.~\eqref{eq:app_density_first_order} reads $m\beta_0\omega^2 q\,e^{-as}\big[\xi(s) - 2a I(s)\big]$. Since $\dot{I} = -aI + \xi$, one has $e^{-as}\big[\xi(s) - 2aI(s)\big] = \frac{\mathrm{d}}{\mathrm{d}s}\big[e^{-as}I(s)\big]$, with $I(0) = 0$. Truncating Eq.~\eqref{eq:app_density_first_order} at first order, we define the linear response density
\begin{equation}
\bar{f}_t(\tilde{\Gamma}) = \tilde{f}^{(0)}_t(\tilde{\Gamma})\Big[1 - \sigma\, m\beta_0\omega^2\, q\, e^{at}I(-t)\Big] .
\label{eq:app_barf}
\end{equation}
For an observable $\mathcal{O}$ depending on $q$ only, one thus finds
\begin{equation}
\hat{\mathbb{E}}^{\bar{f}_t}[\mathcal{O}] = \hat{\mathbb{E}}^{\tilde{f}^{(0)}_t}[\mathcal{O}]  - \sigma\, m\beta_0\omega^2\, e^{at}\int \mathcal{O}(q)\, q\, I(-t)\,\tilde{f}^{(0)}_t(\tilde{\Gamma})\,\mathrm{d}\tilde{\Gamma},
\label{eq:app_barf_expectation}
\end{equation}
with $\mathrm{d}\tilde{\Gamma} = \mathrm{d}q\,\mathrm{d}\theta\,\mathrm{d}\phi$. In the second term, the zeroth-order density factorizes as $\tilde{f}^{(0)}_t(q,\theta,\phi) = p_t(q)\,g_0(\theta,\phi)$, with $p_t(q) = e^{at}f_0(q\,e^{at})$, while $\mathcal{O}(q)\,q$ depends only on $q$ and $I(-t)$ only on the auxiliary variables. Hence
\begin{equation*}
\int \mathcal{O}(q)\, q\, I(-t)\,\tilde{f}^{(0)}_t(\tilde{\Gamma})\,\mathrm{d}\tilde{\Gamma} = \left(\int \mathcal{O}(q)\, q\, p_t(q)\,\mathrm{d}q\right)\hat{\mathbb{E}}^{g_0}\big[I(-t)\big],
\end{equation*}
and
\begin{equation*}
\hat{\mathbb{E}}^{g_0}\big[I(-t)\big] = \int_0^{-t} e^{-a(-t-\tau)}\,\hat{\mathbb{E}}^{g_0}\big[\xi(\tau)\big]\,\mathrm{d}\tau = 0,
\end{equation*}
because each mode of the expansion~\eqref{eq:xi_psi} averages to zero over the uniform phase distribution, for every $\tau$ and every realization of the amplitudes. Therefore,
\begin{equation}
\hat{\mathbb{E}}^{\bar{f}_t}[\mathcal{O}] = \hat{\mathbb{E}}^{\tilde{f}^{(0)}_t}[\mathcal{O}] = \mathbb{E}^{f_0}[\mathcal{O}] + \int_0^t \hat{\mathbb{E}}^{\tilde{f}_0}\!\left[\tilde{\Omega}_0\big(\mathcal{O}\circ\tilde{S}^s_0\big)\right]\mathrm{d}s,
\label{eq:app_LRT_unperturbed}
\end{equation}
the second equality being the exact response formula~\eqref{eq:ERTst} for the unperturbed dynamics.
The same formula holds for the double expectation, upon averaging over the amplitudes. Specifically, for $\mathcal{O} = q^2$, the zeroth-order density~\eqref{eq:app_density_zero_liouville} yields $\mathbb{E}^{\tilde{f}^{(0)}_t}[q^2] = e^{-2at}/(m\beta_0\omega^2)$, which recovers Eq.~\eqref{eq:lin_response_result}.

We now show that Eq.~\eqref{eq:app_LRT_unperturbed} also follows from the expansion of the exact response formula, which clarifies its relation with the classical linear response formula.
For an observable $\mathcal{O}$ depending on $q$ only, and for a fixed realization of the amplitudes, Eq.~\eqref{eq:ERTst} reads
\begin{equation}
\hat{\mathbb{E}}^{\tilde{f}_t}[\mathcal{O}] = \mathbb{E}^{f_0}[\mathcal{O}] + \int_0^t \hat{\mathbb{E}}^{\tilde{f}_0}\!\left[\tilde{\Omega}^{\tilde{f}_0}\big(\mathcal{O}\circ\tilde{S}^s\big)\right]\mathrm{d}s.
\label{eq:app_ERT}
\end{equation}
By Eq.~\eqref{eq:app_traj}, for continuous dynamics and fixed time $s$, the observable along 
the perturbed flow expands as \begin{equation}
\mathcal{O}\circ\tilde{S}^s = \mathcal{O}\circ\tilde{S}^s_0 + \sigma\,\mathcal{O}'\big(q_0 e^{-as}\big)\,I(s) + O(\sigma^2),
\label{eq:app_O_expansion}
\end{equation}
where $I(s)$, $s \ge 0$, is now evaluated along the forward auxiliary trajectory starting from $(\theta_0,\phi_0)$. Multiplying by $\tilde{\Omega}^{\tilde{f}_0} = \tilde{\Omega}_0 + \sigma\,\tilde{\Omega}_1$ and discarding terms of order $\sigma^2$, we obtain the first-order expansion of the exact response:
\begin{equation}
\begin{split}
\hat{\mathbb{E}}^{\tilde{f}_t}[\mathcal{O}] &= \left(\mathbb{E}^{f_0}[\mathcal{O}] +\sigma\int_0^t \hat{\mathbb{E}}^{\tilde{f}_0}\!\left[\tilde{\Omega}_1\big(\mathcal{O}\circ\tilde{S}^s_0\big)\right]\mathrm{d}s\right) \\
&\quad + \left(\int_0^t \hat{\mathbb{E}}^{\tilde{f}_0}\!\left[\tilde{\Omega}_0\big(\mathcal{O}\circ\tilde{S}^s_0\big)\right]\mathrm{d}s + \sigma\int_0^t \hat{\mathbb{E}}^{\tilde{f}_0}\!\left[\tilde{\Omega}_0\,\mathcal{O}'\big(q_0 e^{-as}\big)\,I(s)\right]\mathrm{d}s\right) + O(\sigma^2).
\end{split}
\label{eq:app_ERT_first_order}
\end{equation}
The first two terms on the right-hand side reproduce the usual linear response formula~\eqref{linresponse}, in which the observable is propagated along the unperturbed flow and $f_0$ is invariant under such dynamics, so that $\tilde{\Omega}_0 \equiv 0$ and $\sigma\tilde{\Omega}_1$ plays the role of the field-dependent dissipation function. The remaining two terms vanish identically in that case, and are generally nonzero when, as in the present model, $f_0$ is not invariant under the unperturbed dynamics: the third term accounts for the unperturbed evolution of $f_0$, while the fourth term is a first-order correction due to the deviation $\sigma I$ of the trajectory from its unperturbed realization.

The linear response prediction, defined as the first-order expansion of the exact response, comprises all four terms of Eq.~\eqref{eq:app_ERT_first_order}. In the present model, however, both first-order terms vanish for every observable $\mathcal{O}$ depending on $q$ only. Indeed, since $q_0$ is independent of the auxiliary variables, they factorize as
\begin{equation*}
\hat{\mathbb{E}}^{\tilde{f}_0}\!\left[\tilde{\Omega}_1\big(\mathcal{O}\circ\tilde{S}^s_0\big)\right] = \omega^2 m\beta_0\,\mathbb{E}^{f_0}\!\left[q_0\,\mathcal{O}\big(q_0 e^{-as}\big)\right]\hat{\mathbb{E}}^{g_0}[\xi_0]
\end{equation*}
and
\begin{equation*}
\hat{\mathbb{E}}^{\tilde{f}_0}\!\left[\tilde{\Omega}_0\,\mathcal{O}'\big(q_0 e^{-as}\big)\,I(s)\right] = \mathbb{E}^{f_0}\!\left[\tilde{\Omega}_0\,\mathcal{O}'\big(q_0 e^{-as}\big)\right]\hat{\mathbb{E}}^{g_0}[I(s)],
\end{equation*}
and both $\hat{\mathbb{E}}^{g_0}[\xi_0]$ and $\hat{\mathbb{E}}^{g_0}[I(s)]$ vanish, each mode of the expansion~\eqref{eq:xi_psi} averaging to zero over the uniform phase distribution. The vanishing of the fourth term implies that the linear response prediction takes the form of Eq.~\eqref{eq:LRT_formula}, with $q^2$ replaced by $\mathcal{O}$, in which the observable is propagated along the unperturbed flow; the vanishing of the second one implies, moreover, that it contains no contribution of the noise, and reduces to the evolution of the initial density under the unperturbed dynamics $\dot{q} = -aq$, i.e.\ with the noise switched off, Eq.~\eqref{eq:app_LRT_unperturbed}.

We conclude this appendix with a criterion for the range of validity of the linear response formula. We square Eq.~\eqref{eq:app_traj} and average over $q_0$, the phase and the amplitudes. In the large-$T$ limit, where the noise correlation reduces to $\delta(\tau-\tau')$ (see Prop.~\ref{prop:prop1}), one obtains
\begin{equation*}
\mathbb{E}^{f_t}[q^2] = e^{-2at}\,\mathbb{E}^{f_0}[q_0^2] + \sigma^2\!\int_0^t e^{-2a(t-\tau)}\,\mathrm{d}\tau .
\end{equation*}
The first term on the right-hand side is the initial variance contracted by the deterministic relaxation, which, by Eq.~\eqref{eq:app_LRT_unperturbed}, is the linear response prediction,
\begin{equation}
e^{-2at}\,\mathbb{E}^{f_0}[q_0^2] = \mathbb{E}^{\bar{f}_t}[q^2] = \frac{e^{-2at}}{m\beta_0\omega^2},
\label{eq:app_var_initial}
\end{equation}
whereas the second is the variance accumulated by the noise, which, using $\sigma^2 = 2/(m\beta\gamma)$, reads
\begin{equation}
\sigma^2\!\int_0^t e^{-2a(t-\tau)}\,\mathrm{d}\tau = \frac{\sigma^2}{2a}\big(1-e^{-2at}\big) = \frac{1-e^{-2at}}{m\beta\omega^2},
\label{eq:app_var_noise}
\end{equation}
and coincides with $\Delta\mathbb{E}[q^2]$ in Eq.~\eqref{eq:Delta}. The criterion requires that the variance~\eqref{eq:app_var_noise} be small compared with~\eqref{eq:app_var_initial}, i.e.\ $(m\beta_0\gamma\sigma^2/2)\,(e^{2at}-1) \ll 1$. For $\beta$ either larger or smaller than $\beta_0$, this condition holds for $t$ well below
\begin{equation*}
t^* = \tau_r\ln\!\Big(1 + \frac{2}{m\beta_0\gamma\sigma^2}\Big) = \tau_r\ln\!\Big(1 + \frac{\beta}{\beta_0}\Big),
\end{equation*}
where $\tau_r = (2a)^{-1}$ is the relaxation time of the second moment introduced in Section~\ref{sec:LRT}. For $\beta<\beta_0$ one has $t^* < \tau_r\ln 2\simeq 0.69\, \tau_r$, while for $\beta_0/\beta \ll 1$, $t^* \simeq \tau_r\ln(\beta/\beta_0) \gg \tau_r$, which nevertheless remains finite.

\bibliographystyle{abbrv}
\bibliography{bibliografia_lavoro1}
\end{document}